\documentclass{pas}

\usepackage[utf8]{inputenc}
\usepackage{amsmath}

\usepackage[nopatch]{microtype}
\usepackage{booktabs}
\usepackage{ae,aecompl}
\usepackage{aas-macros}

\usepackage[]{lineno}
\usepackage{graphicx}	% Including figures
\usepackage{amssymb}	% Extra maths symbols
\usepackage{ae,aecompl}

\DeclareUnicodeCharacter{2009}{ } % Duct-Tape solution to missing unicode
\usepackage{hyperref}
\hypersetup{
	colorlinks=true,
	urlcolor=blue,    % color of external links
	linkcolor=blue,  % color of toc, list of figs etc.
	citecolor=blue,   % color of links to bibliography
}

\usepackage{orcidlink}

\usepackage{graphicx}	% Including figure files
\usepackage{algorithm,algcompatible}
\usepackage{braket}
\usepackage{float, makecell, array}
\usepackage{xspace, bm}
\usepackage{pdflscape}
\usepackage{longtable, makecell}
\usepackage{subfigure}
\usepackage{multirow, multicol}
\usepackage{fontawesome} % for GitHub icon (from Font Awesome)

\usepackage{multirow}
\usepackage{makecell}

\newcommand{\hbeta}{H$\beta$\xspace}
\newcommand{\mgii}{Mg\textsc{ii}\xspace}
\newcommand{\civ}{C\textsc{iv}\xspace}

\newcommand{\beq}{\begin{equation}}
\newcommand{\eeq}{\end{equation}}
\newcommand{\bea}{\begin{eqnarray}}
\newcommand{\eea}{\end{eqnarray}}
\newcommand{\multilinecomment}[1]{}
\newcommand{\logof}[1]{\ln \left|  #1 \right|}
\newcommand{\qm}[1]{``#1''}
\newcommand{\sqm}[1]{`#1'}
\newcommand{\hescan}{Laplace Quadrature\xspace}
\newcommand{\sviscan}{SVI Quadrature\xspace}

\newcommand{\dayu}{\mathrm{d}}
\newcommand{\ozdes}{OzDES\xspace}
\newcommand{\sdss}{SDSS\xspace}

\newcommand{\javelin}{\texttt{JAVELIN}\xspace}
\newcommand{\emcee}{\texttt{emcee}\xspace}
\newcommand{\PyCCF}{\texttt{PyCCF}\xspace}
\newcommand{\pyroa}{\texttt{PyROA}\xspace}

\newcommand{\cream}{\texttt{CREAM}\xspace}
\newcommand{\mica}{\texttt{MICA}\xspace}
\newcommand{\litmus}{\texttt{LITMUS}\xspace}
\newcommand{\python}{\texttt{python}\xspace}

\newcommand{\numpyro}{\texttt{numpyro}\xspace}
\newcommand{\jax}{\texttt{jax}\xspace}
\newcommand{\jaxopt}{\texttt{jaxopt}\xspace}
\newcommand{\jaxns}{\texttt{JAXNS}\xspace}
\newcommand{\tinyGP}{\texttt{tinyGP}\xspace}
\begin{document}

\lefttitle{Publications of the Astronomical Society of Australia}
\righttitle{Hugh McDougall}

\jnlPage{1}{17}
\jnlDoiYr{2026, Vol 43-e108}
\doival{10.1017/pasa.2026.10149, DES-2025-860 FERMILAB-PUB-25-0824-PPD}

\articletitt{Research Paper}

\title{\litmus: Bayesian Lag Recovery in Reverberation Mapping with Fast Differentiable Models}

\author{
\gn{Hugh} \sn{McDougall}\orcidlink{0009-0008-5846-1543}$^{1}$, \gn{Benjamin~J.~S. } \sn{Pope}\orcidlink{0000-0003-2595-9114}$^{2,1}$, \gn{Tamara M.} \sn{Davis}\orcidlink{0000-0002-4213-8783}$^{1}$
}

\affil{
$^{1}$School of Mathematics and Physics, University of Queensland, St Lucia, QLD 4072, Australia, $^{2}$ School of Mathematical \& Physical Sciences, 12 Wally's Walk, Macquarie University, Macquarie Park, NSW 2113
}

\corresp{Hugh McDougall, Email: hughmcdougallemail@gmail.com}

\citeauth{McDougall et al., litmus: Bayesian Lag Recovery in Multi-Reverberation Mapping with Fast Differentiable Models {\it Publications of the Astronomical Society of Australia} {\bf 43}, e108. https://doi.org/10.1017/pasa.2026.10149}

\history{(Received 24 May 2025; revised 18 December 2025; accepted 7 January 2026))}

\begin{abstract}
Reverberation mapping is a technique in which the mass of a Seyfert I galaxy's central supermassive black hole is estimated, along with the system's physical scale, from the timescale at which variations in brightness propagate through the galactic nucleus. This mapping allows for a long baseline of time measurements to extract spatial information beyond the angular resolution of our telescopes, and is the main means of constraining supermassive black hole masses at high redshift. The most recent generation of multi-year reverberation mapping campaigns for large numbers of active galactic nuclei (AGN) (e.g. \ozdes) have had to deal with persistent complications of identifying false positives, such as those arising from aliasing due to seasonal gaps in time-series data. We introduce \litmus (Lag Inference Through the Mixed Use of Samplers), a modern lag recovery tool built on the \qm{damped random walk} model of quasar variability, built in the automatic differentiation framework \jax. \litmus is purpose-built to handle the multimodal aliasing of seasonal observation windows and provides Bayesian evidence integrals for model comparison and null hypothesis testing, a more quantified alternative to existing post-fit selection methods. \litmus also offers a flexible and modular framework for using more expressive high dimensional models for the AGN variability, and includes \jax-enabled implementations of other popular lag recovery methods like nested sampling and the interpolated cross correlation function. We test \litmus on a number of mock light curves modelled after the \ozdes sample and find that it recovers their lags with high precision and successfully identifies spurious lag recoveries, reducing its false positive rate to drastically outperform the state of the art program \javelin. \litmus's high performance is accomplished by an algorithm for mapping the Bayesian posterior density which both constrains the lag and provides Bayesian evidences for model comparison and null hypothesis testing while outperforming nested sampling in computational cost by an order of magnitude. \href{https://github.com/HughMcDougall/litmus/}{\faGithub} 
\end{abstract}

\begin{keywords}
galaxies: active – galaxies: nuclei – quasars: general  – methods: data analysis – gravitational lensing: strong
\end{keywords}

\maketitle

%%%%%%%%%%%%%%%%%%%%%%%%%%%%%%%%%%%%%%%%%%%%%%%%%%

%%%%%%%%%%%%%%%%% BODY OF PAPER %%%%%%%%%%%%%%%%%%

\section{Introduction}
\label{sec: introduction}

Reverberation mapping (RM) is a technique in which we measure the way that fluctuations of a source's brightness propagate through its physical components, using the delays in the signal stimulated by this flux to estimate the physical scale of the system. In this way, RM substitutes temporal resolution for angular resolution in our observations, allowing us to tease out spatial information about in-sky point-sources. The technique was first introduced \citep{Blandford_McKee_1982, Peterson_1993} in measuring the scale of the broad line region (BLR) of active galactic nuclei (AGN), measuring the lag between variations in the broad-band photometry, dominated by the AGN's central accretion disk, and the emission lines excited in the BLR, observed via spectroscopy. Assuming a virialized orbit of the BLR, this radius (together with velocity measurements from Doppler broadening of the emission lines), allows us to measure the mass of the AGN's central supermassive black hole. 

Though uncertainty in the geometry and kinematics of the system mean that BLR RM requires a low-redshift anchor to measure masses, it nevertheless forms a powerful tool for measuring quasar masses to cosmological distances. Recent years have seen the close of the first generation \qm{industrial-scale} RM surveys, namely the Sloan Digital Sky Survey \citep[\sdss;][]{SDSS-Shen_2015} and the Australian Dark Energy Survey \citep[\ozdes;][]{King_2015,OZDES-DR2-Lidman_2020}, with large numbers of AGN observed out to high redshifts. Such surveys measure AGN masses directly, but also use their measurements to constrain the \qm{radius-luminosity}, or $R-L$, scaling relationship for AGN: an observed power-law between the scale of the BLR and the  mean photometric luminosity of the quasar \citep{Kaspi_2000}. 

Though these large scale surveys collect hundreds to thousands of RM lags, they are impacted by the problem of aliasing: the emergence of multiple peaks in the lag posterior distribution that emerges from the annual six month seasonal gaps in observations. Aliasing has proven to be a major issue in long-baseline RM, with significant efforts being directed towards either suppressing spurious lags in the posterior distribution \citep[e.g.][]{SDSS-Grier_2017, CIV_Grier_2019} or filtering out suspect lags through quality cuts \citep[e.g.][]{OzDES-Penton_2021, OzDES-Yu_2023,SDSS-Shen_2023, OzDES-Penton_2025}. Such methods often disagree about how many sources to keep, and the most stringent of them can lead to as many as $90\%$ of sources being discarded to ensure the purity of the remaining $10\%$ \citep[see][for a comparison and discussion]{OzDES-McDougall_2025}.

As well as extending into higher redshifts and farther sources, RM has also evolved to probe both smaller and larger physical scales of the AGN, with RM being applied to both the large dusty torus that orbits beyond the BLR \citep[e.g.][]{Suganuma_2006, Koshida_2014, Minezaki_2019}, and also to smaller scales (roughly light-days) of the accretion disk itself \citep[][]{Fausnaugh_2016, OzDES-Yu_2020}. RM of the accretion disk typically measures lags between observations in different photometric filters, comparing the bluer light from the hot interior of the disk to the redder outer edge, allowing both the size and temperature profile of the disk to be constrained. Where X-ray observations are available, we can also probe the emission from the black hole's corona that is believed to be the driving signal of the AGN variability \citep{Cackett_2007}. Though RM alone only directly measures scale, it can also be used to infer properties in other dimensions. For example, in recent years intensive broad-band RM has revealed possible evidence of viscous motion of the disk material, identifying a \qm{long} timescale of possibly negative lag associated with the inwards migration of hot material from the disk's cooler outer edge to its hot interior \citep[e.g.][]{Santisteban_2020, Secunda_2023}. Such measurements imply a disk that is significantly thicker than the traditional \qm{optically thick, geometrically thin} disk assumed by the famous thin-disk model of \citet{Shakura_1973}. For a review of these techniques, we direct the reader to the excellent review of RM at different scales by \citet{Cackett_2021}.

High cadence of X-ray reverberation allows for complex Fourier analysis that infers multiple lags from different parts of the signal \citep[for a more thorough review, see][]{Uttley_2014}. A similar increase in complexity allows for \qm{velocity resolved} reverberation mapping, which aims to constrain the geometry of the BLR in addition its characteristic scale. All other RM falls under the roof of time delay estimation, most commonly between two light curves at a time, in which we aim to constrain a single characteristic delay between the two signals. For quasars we have a convenient statistical description of AGN variability: though they lack deterministic light curves, their stochastic variations are found to follow closely to the power spectral density of the \qm{damped random walk} (DRW), a first order continuous auto-regressive process (i.e. one that acts like red noise at short timescales). The DRW is an example of a Gaussian process (GP), a class of stochastic signals that are well suited to modelling AGN variability well in general \citep[we give a brief overview in this paper, but for a full review of GP's in astronomy see][]{Aigrain_2023_GPReview}.

Since its inception, RM has seen a steadily evolving bank of tools for this time delay estimation / lag recovery task. These broadly fall into two categories: those that model the AGN light curves as a Gaussian Process and are  agnostic about the underlying statistical behaviour of the light curve and parameters, and those that use our understanding of the statistical properties of the AGN's stochastic variations to fit lags with a Bayesian generative model.

Despite its conceptual simplicity, this is far from a trivial task. Aside from the ever growing tool-chest of AGN RM, the adjacent field of estimating quasar time delays from gravitational lensing, essentially the same statistical problem, encountered enough issues to warrant the Time Delay Lens Modelling Challenge \citep{ding_2018_TDLMC1, Ding_2021_TDLMC2}, an open competition amongst a number of teams to find even a single robust and reliable method for inferring lags. Though the more rigorous GP models have been in place for well over a decade, non-GP methods are still often used thanks to their low numerical cost and apparent lower sensitivity to aliasing. In this paper, we show that the aliasing problem of the GP methods is in fact overstated, and a result not of the fundamental statistical properties of RM or seasonal observations but rather a numerical artefact owing to a choice of sampling method incorrect for this problem. In this paper we provide a new code, \litmus, which solves many of the challenges of aliasing while putting rigorous statistical constraints on its impact, discussed in detail in Section~\ref{sec: litmus}. We describe the statistical modelling and numerical fitting of this new method, and demonstrate that this approach can both recover lags in mock data better than existing tools and that it can distinguish between true lags and spurious false positives.

\section{Principles of Reverberation Mapping \& Complications Therein}
\label{sec: RM_theory}

In the most general sense, RM is a technique of measuring the radial scale of a system, however in the context of surveys like \ozdes the goal is to constrain the mass of the central black hole via the size of the BLR. This is done by simplifying the BLR to be a thin disk with a single characteristic radius/ reverberation lag $R=c\cdot\Delta t$, where $c$ is the speed of light. Assuming the BLR is in a virialized orbit about the SMBH such, the virial mass can be estimated with a line-of-sight velocity dispersion from the doppler broadening of the emission line profile. The unknown geometry and kinematics are captured in the \qm{virial factor}, calibrated with local anchors of known mass \citep[e.g. from the $M_\mathrm{SMBH}-M_\mathrm{*}$ relationship;][]{Woo_2015, Grier_2013}. This factor has high population dispersion, imposing on RM mass measurements an uncertainty floor of roughly a quarter. RM lags are also used to constrain the power-law scaling relationship between BLR radius and AGN luminosity (the $R-L$ relationship). Accurate measurement of lags is crucial not only for inferring AGN masses, but also for constraining the slope, offset and scatter parameters in this relationship \citep[e.g.][]{OzDES-McDougall_2025}.

There has been a broad spectrum of approaches in attempting to consistently and reliably measure $\Delta t$ across all AGN (see Section~\ref{sec: existing_methods}), and for \litmus we adopt the  approach of Bayesian forward modelling, i.e. one in which the AGNs signals are characterised by some set of model parameters $\theta$ which we have constrained to some prior probability distribution $\pi(\theta)$, and for which any $\theta$ has a likelihood $\mathcal{L}(D \vert \theta)$ of reproducing our observational data $D$ such that there is a joint distribution $P(\theta \vert D)$\footnote{In this paper, we use  $P(\theta)$ for the normalised probability distribution, and $\mathcal{P}(\theta)$ to represent the un-normalised joint distribution, i.e. $P(\theta) = \frac{1}{Z} \mathcal{P}(\theta\vert)$, where $Z$ is the model evidence per Equation~\ref{eq: evidence_definition}.} for parameters and observations:
\begin{equation}
    \mathcal{P}(\theta \vert D) = \mathcal{L}(D \vert \theta)\pi(\theta)=Z \times \mathcal{P}(\theta)
    .
    \label{eq: joint_definition}
\end{equation}
We describe a few existing methods for recovering lags within this Bayesian GP framework in Section~\ref{sec: existing_methods-parametric}. This is the approach taken by \litmus. Unique to \litmus is the calculation of Bayesian model evidence, the total \qm{probability mass} of the posterior distribution:
\begin{equation}
    Z = \int \mathcal{P}(\theta \vert D) d\theta
    .
    \label{eq: evidence_definition}
\end{equation}
This evidence is a total \qm{goodness of fit} for the model to the data, and comparing the evidence for two models (their ratio being the \qm{Bayes Factor}) is a means of comparing which model is more supported by observation. This is the approach taken by \litmus in determining whether lag recoveries are statistically significant.

As our understanding of the AGN variability has evolved we have come to describe their stochasticity as a GP, wherein the time series observations $y$ are Gaussian-correlated observations, such that the model likelihood is:
\begin{equation}
    \mathcal{L}(\theta \vert D) = \frac{1}{\sqrt{(2\pi)^N\det(C)}} 
    \exp\left( -\frac{1}{2}y^TC^{-1}y \right)
    ,
    \label{eq: GP_likelihood}
\end{equation}
where $C$ is the \sqm{covariance matrix} constructed for all observations over all light curves, and $y$ is the vector of observations after subtracting off the signal means (these means also being model parameters). This includes both the intrinsic variance / covariance of the GP in the matrix $S$, and the approximately uncorrelated white noise measurement uncertainty for each observation ($E_i$ for measurement $i$) in the diagonal matrix $N$:
\begin{equation}
    C = S + N, \; N=\delta_{ij} E_i E_j
\end{equation}

If the underlying continuum light curve is described by a GP with a covariance function:
\begin{equation}
    \phi_c(t) = \braket{y_c(t-t'),y_c(t)}
    ,
    \label{eq: cc_covar}
\end{equation}
and some the response function is \qm{shifted, scaled and blurred}, i.e. multiplied by some amplitude, mean offset, mean lag and then blurred by smoothing kernel $\psi(t)$, then the auto-covariance of the response signal is:
\begin{equation}
    \phi_r(t)
    = \int_{-\infty}^{\infty}{\phi_c(t)\psi(t')}dt'
    ,
\end{equation}
and the covariance between the response signal and continuum is:
\begin{equation}
    \phi_{rc}(t)
    = \iint_{-\infty}^{\infty}{\phi_c(t)\psi(t')\psi(t'')}dt'dt''
    .
\end{equation}

\begin{figure*}
    \centering
    \includegraphics[width=0.9\linewidth]{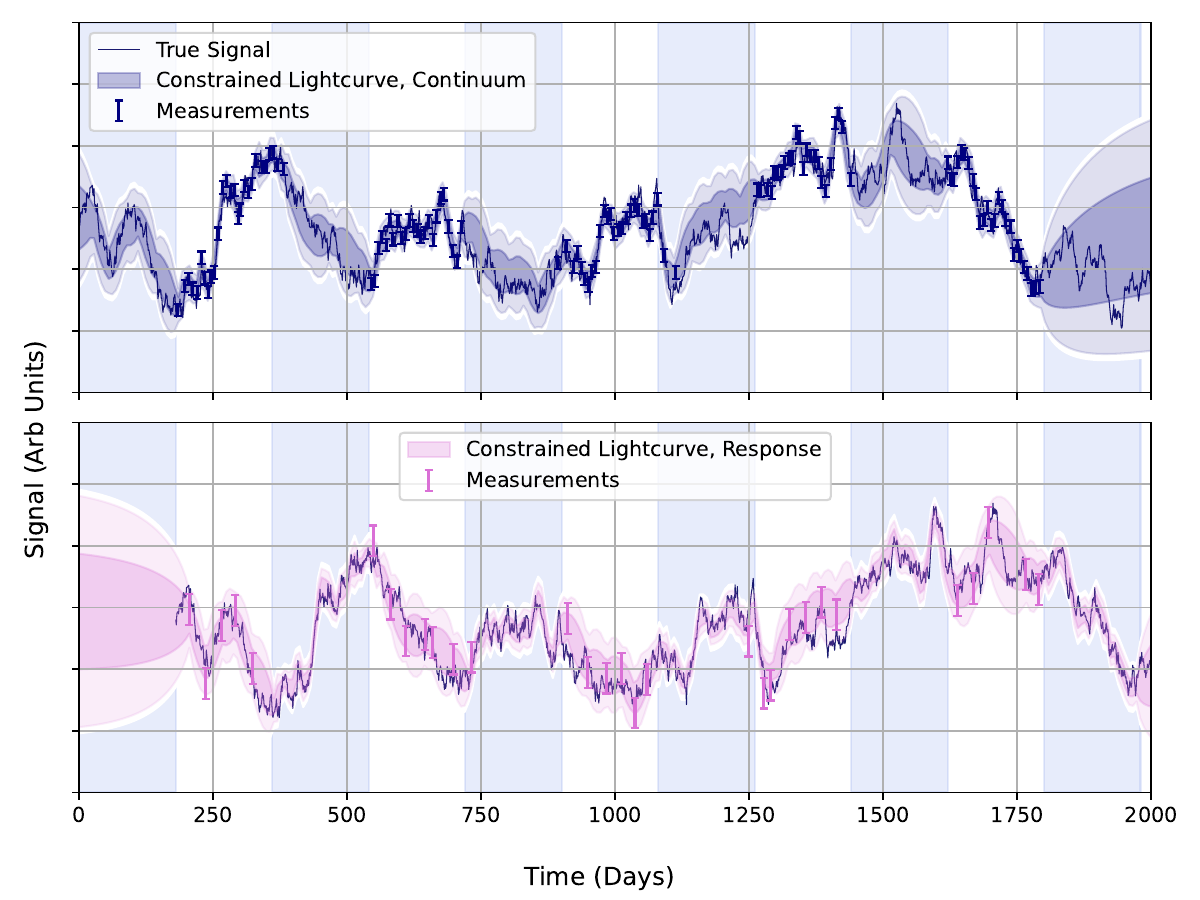}
    \caption{A demonstration of the sort of light curves that GP modelling can reconstruct from observations. For some time-series observations (error bars) a particular GP models the entire family of underlying light curves that exhibit the power spectral density of the GP, conditioned on how well they fit the observations. In this example the light curves is fit as a DRW with $\tau=200 \dayu$ and $\sigma=1$, both in arbitrary units for this demonstrative example. The shaded regions represent the $1$ and $2 \sigma$ contours of the distribution of all such walks.}
    \label{fig: GP_example}
\end{figure*}

\noindent If the parameters are known, this covariance function can also be used to reconstruct confidence intervals for the behaviour of the light curve between observations (e.g. Figure~\ref{fig: GP_example}).

It has been empirically observed that AGN variability matches closely to the damped random walk \citep[DRW][]{Kelly_2009, Kozlowski_2010a, MacLeod_2010}, in which covariance function obeys the double exponential Laplace distribution:
\begin{equation}
    \phi_c(t)=\sigma_c^2 \exp \left( \frac{-\lvert t \rvert}{\tau} \right)
    .
    \label{eq: DRW_covar}
\end{equation}
Though actual AGN variability differs slightly from a DRW \citep{Zu_2013}, it has been found that the exact choice of GP / covariance function has little impact on lag recovery \citep{OZDES-Yu_2019}.

Although \litmus works only with single lags / response signals in its initial release, this model also extends to the multi-lag case by having response-response covariance for two response signals with transfer functions $\psi_1(t)$ and $\psi_2(t)$:
\begin{equation}
    \phi_{rr}(t)
    = \iint_{-\infty}^{\infty}{\phi(t)\psi_1(t')\psi_2(t'')}dt'dt''
    .
\end{equation}
This multi-lag RM sees considerable use in mapping the temperature profile of accretion disks \citep[e.g.][]{OzDES-Yu_2020}, where different colour temperatures occur at different annular radii.

\subsection{The Aliasing Problem}
\label{sec: RM_theory-aliasing}

As reverberation mapping first pressed into the regime of industrial scale RM, the endeavour soon ran afoul of the confounding effects of \sqm{aliasing}: a suite of problems emerging from the combination of low precision measurements (shorter exposures, looser cadence) and the half-yearly seasonal gaps in the observations. Aliasing manifests as a tendency for lag recovery methods to over-report lags at and around the \qm{aliasing peaks}, which correspond to minimal data overlap between the continuum and response curves ($\approx 180 \;\dayu$, $540 \;\dayu$ etc.). This aliasing problem is a major confounding factor in modern RM, with the arising false positive lags dominating the mass measurements and $R-L$ constraints when false positives are not screened.

The traditional approach to the aliasing problem has been to take a frequentist approach of quality control: drawing a statistical cordon around the entire lag recovery method, inclusive of the statistical model, numerical model and lag recovery software within a single \sqm{black box} and determining some fit quality measure for determining how likely a recovery is to be erroneous. The \ozdes team has adopted a consistent, if steadily evolving, framework for this approach: using \javelin as its primary lag recovery method, in conjunction with \PyCCF as a validation method, and using simulation-based measures of the the likelihood of a false positive to winnow down to a high purity final data set with $<10\%$ contamination by spurious lag recoveries. Across its various data releases \sdss have trialled multiple different anti-aliasing regimes. In earlier releases 
\citep{SDSS-Grier_2017, CIV_Grier_2019}, they use \cream \citep{Starkey_2015_CREAM} for their lag posterior and validate through a combination of down-weighting lags that lie near aliasing peaks and rejecting recoveries that allow for non-physical negative lags. In their final data release they adopt a less stringent selection criteria, using \pyroa \citep{pyroa-ferus_2021} as their primary lag recovery method and testing for significance with the $r^2$ recovered by \PyCCF. For a full review and comparison of these selection criteria, we direct the reader to the \ozdes RM wrap-up paper \citep{OzDES-McDougall_2025}.

\begin{figure*}
    \centering
    \includegraphics[width=0.9\linewidth]{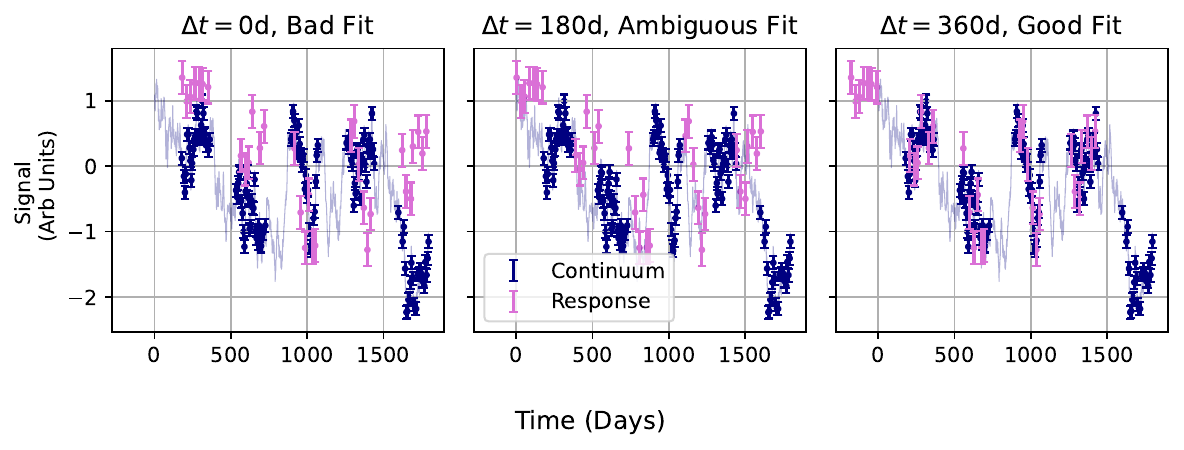} \\
    \includegraphics[width=0.9\linewidth]{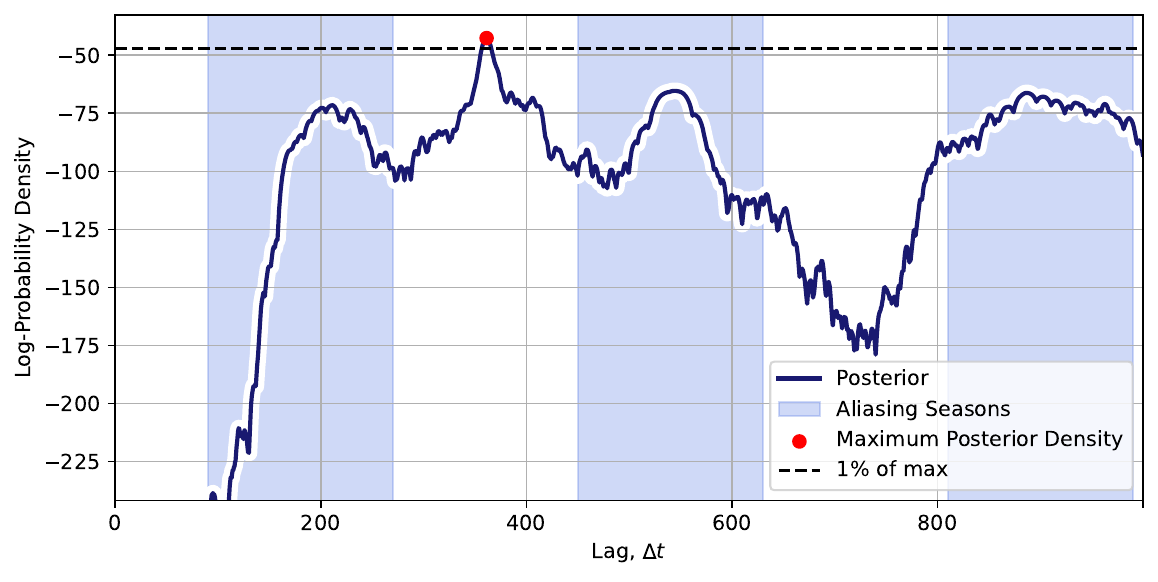}
    \caption{A demonstration of the source of the aliasing problem, specifically in the context of a parametric GP model. Top shows mock data with cadence, measurement uncertainty and baseline similar to \ozdes with a DRW timescale of $\tau=200 \dayu$ and a true lag of $\Delta t=360 \dayu$. From left to right the sub-panels show lags being tested at $\Delta t=0 \dayu$, $180 \dayu$ and $360 \dayu$. The left panel is clearly a bad fit as near simultaneous observations are in clear tension, and the right panel is a clear good fit as we see very little tension. The middle panel, corresponding to the first aliasing peak, is an ambiguously good fit; the lack of overlap means we cannot observe clear tensions between the light curves. The bottom panel shows the (un-normalised) log-natural of the posterior distribution, with all non-lag parameters fixed at their true values. At \qm{on-season} lags (un-shaded) we can easily reject bad fits, and so the posterior is extremely low. During the off-season lags (blue shading) there are local optima arising from the ambiguity. The mode associated with the true lag (red dot) is clearly defined and dominates over aliasing modes, with the rest of the posterior being $<1 \%$ of the maximum posterior density in this well behaved, high SNR example. Even so, the posterior still suffers from the rough geometry and multimodality that introduces numerical challenges in navigating it.}
    \label{fig: aliasing_explained}
\end{figure*}

In the abstract sense, aliasing occurs because testing the half-yearly \qm{off-season} lags coincides with little to no overlap in data between light curves. Lag recovery is as much about rejecting poor fits as identifying good ones, and such a lack of overlap means we have fewer chances to identify tensions that indicate such bad fits. Between seasonal gaps, over-loose interpolations are vague and weekly constraining, while over-generous interpolations are at the mercy of coincidence and can produce spurious structure. In either case, failure-modes of the light curve reconstruction / lag testing are at their worst at these difficult lags.  In Figure~\ref{fig: aliasing_explained} we demonstrate this for mock data, and in particular show the multimodality that occurs in the lag posterior for the specific case of a parametric GP model. Aliasing becomes worse when the true lag is near an aliasing peak, as these are more ambiguously observed and produce  a shallower peak in the posterior. In such cases we can only observe tension / discontinuity in the reconstructed light curve in the joins between observation seasons, and so aliasing becomes more pronounced for more rapid timescales of variability (smaller $\tau$ in Equation~\ref{eq: DRW_covar}) and for fewer observed seasons. 

\begin{figure}
    \centering
    \includegraphics[trim={0 1.4cm 0 0}, clip, ]{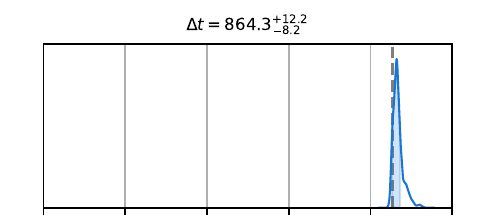}
    \includegraphics[trim={0 0 0 0.2cm}, clip, ]{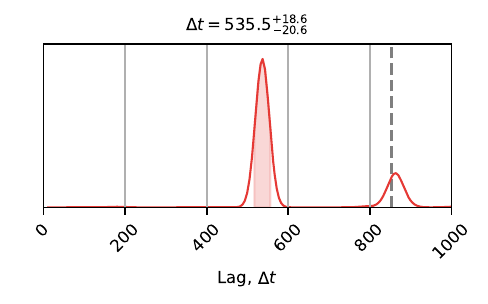}\\
    \caption{A demonstration of the failure mode of the Affine-Invariant Ensemble Sampler (AIES), the MCMC proposal algorithm used by \emcee, in multi-modal distributions. Both top and bottom panels are posterior distributions generated from the same mock data with a true lag at $\Delta t = 854 \dayu$ (dashed line), with the bottom panel being the result from the AIES, the same MCMC sampler as \javelin, while the top is found from exhaustive sampling of the prior range. The AIES estimate for the posterior has produced an aliasing peak at $\Delta t = 540 \dayu$ where none truly exists due to its ensemble of live sampling points becoming pinned at this minor mode.}
    \label{fig: emcee_multimodal_failure}
\end{figure}

In the example shown in Figure~\ref{fig: aliasing_explained}, the lag is well constrained to its true value as the aliasing peaks are several orders of magnitude shallower. However some Bayesian modelling tools can fail even in these cases due to the secondary numerical challenges that aliasing poses. \javelin fits lags by sampling the posterior with \emcee, an implementation of the Affine Invariant Ensemble Sampler \citep{Goodman_2010_AIES}. Though \emcee has a well earned reputation as a robust and reliable sampler, it is decidedly not suited to such multimodal distributions. Its proposal method, which involves drawing a chord between two samples and performing a \qm{stretch-move} to propose a new sample, cannot easily mix between well separated modes. The result is the ensemble of live points becomes partially \sqm{pinned} at the aliasing modes, over-sampling them and giving the false appearance of a significant bulk of posterior density, sometimes multiple orders of magnitude over the true height of the peak. In Figure~\ref{fig: emcee_multimodal_failure} we see the dangerous failure mode that can result: the mirage appearance of an aliasing peak in the reported posterior where none truly exists. 

In addition to aliasing, there is a less discussed but still important matter of the sharp dips and valleys in the posterior that give is a \qm{rough} geometry, particularly in the low log-density regions (more commonly around the on-season lags, between the aliasing peaks). The existence of these \qm{furrows} does not obscure the statistical results, but it can make the posterior extremely difficult to navigate for some algorithms. The sharp gradients and potential energy \qm{walls} can cause samplers, particularly gradient based samplers like Hamiltonian Monte Carlo \citep{Duane_1987_HMC01, Neal_1996_HMC02,Betancourt_2018_HMC03} to become stuck, and the cases where these dips go to extreme negative values can lead to unstable computational overflow. These furrows arise from the same source as aliasing, but on a shorter timescale. At lags where two individual observations overlap between continuum and response, the sharp tension between them causes a massive penalty to the log-likelihood, forming a deep valley in the log-posterior. Furrows are then more severe for smaller measurement uncertainty and higher measurement cadence, paradoxically becoming more of an issue the better our measurements are. For signals with slower  variations (larger $\tau$ in Equation~\ref{eq: DRW_covar}), the longer correlation timescale smooths this effect out somewhat.

\section{Existing Methods}
\label{sec: existing_methods}

Reverberation mapping has progressed markedly over its decades long history, both in our understanding of the physics and of the best-practice in constraining the lags. Lag recovery methods can be broadly grouped into two categories: the Bayesian GP methods which make use of our understanding of the AGN signal statistics, and the non-GP methods which aim to be as agnostic as possible about the signal, using only flexible means of interpolation and measures of goodness of fit.

\litmus's main fitting methods belong to the GP-based class, and so in Section~\ref{sec: existing_methods-parametric} we discuss this family and, in examine in particular its widely used exemplar \javelin and its limitations. We also discuss non-GP methods in Section~\ref{sec: existing_methods-nonparametric}, in particular the method used by \PyCCF. In Section~\ref{sec: existing_methods_other} we give a brief overview of methods that are not used by \ozdes / not directly compared against \litmus in this paper.

\subsection{Gaussian Process Methods / \javelin}
\label{sec: existing_methods-parametric}

The main focus of this paper is the class of GP methods that leverage our understanding of AGN signal statistics to work within the framework of a full Bayesian generative model. The core principle of these methods is to take the description of the AGN variability as a Gaussian process (discussed in detail in Section~\ref{sec: RM_theory}), for which there is a closed form and (somewhat) easily evaluable likelihood function. 

Different methods of this class differ in three ways:
\begin{enumerate}
    \item How they construct their covariance matrix (what GP to describe the underlying AGN variability with, how to describe the transfer function of the response).
    \item What statistical methods they use to map the Bayesian posterior distribution.
    \item What parameter-space they define their priors over, i.e. a purely phenomenological set of parameters to describe the signals (signal mean, amplitude etc, in the style of \javelin), or a more physically motivated set of parameters (SBMH mass, accretion rate etc, in the style of \cream).
\end{enumerate}

% \subsubsection{\javelin}

\javelin is the longest standing of the GP fitting methods and has the distinction of being the most widely used. \javelin \citep{JAVELIN-Zu_2010} is a successor to the \texttt{FORTRAN}-based \texttt{SPEAR}. It models the AGN light curves as a damped random walk (following the covariance functions outlined in Section~\ref{sec: RM_theory}) and models the response smoothing with a \qm{top-hat} smoothing function:

\begin{equation}
    \psi(t) = \frac{1}{b}
    \begin{cases}
    1, \left| t  \right| \le \frac{b}{2}\\
    0, \left| t  \right| > \frac{b}{2}
    \end{cases}
    .
    \label{eq: tophat_tfer_func}
\end{equation}
Noting that the exact form of smoothing tends to have little impact on the lag recovery in BLR RM \citep{OZDES-Yu_2019} it is common-place when using \javelin to fix the smoothing scale, $b$, to some fixed value shorter than the observation cadence \citep[e.g.][]{OzDES-Penton_2021}, a choice found to have little impact on lag recovery in \ozdes-like observational surveys \citep{OZDES-Yu_2019}.

\javelin has seen considerable use in the first industrial scale generation of RM, being the \qm{primary} lag recovery method for all of \ozdes \citep{OzDES-Hoormann_2019, OzDES-Malik_2023, OzDES-Penton_2021, OzDES-Yu_2021, OzDES-Yu_2023}, and some of the \sdss releases \citep[e.g.][]{SDSS-Shen_2015}. For constraining its parameters, including the lag, \javelin uses the MCMC package \emcee \citep{emcee-Foreman_Mackey_2013}, though earlier versions used the classic Metropolis-Hastings Algorithm \citep{Metropolis_1953-MetHaste}. As we discuss in Section~\ref{sec: RM_theory-aliasing}, \emcee fails to properly handle the multimodal log-probability distributions of seasonal light curves. This makes it a poor fit to the trials of aliasing, failing to converge in even high signal to noise cases.

Other GP-based lag recovery programs include \cream \citep{starkey_2017_creamerrorscaling} and \mica \citep{Li_2016_MICA}, discussed in more detail in Section~\ref{sec: existing_methods_other}. In this paper we focus our attention on \javelin's AEIS fitting method, as \javelin is the primary \ozdes lag recovery code and has a history as the most prevalent tool for BLR RM.

\subsection{Non-Gaussian Process Methods / \PyCCF}
\label{sec: existing_methods-nonparametric}

GP methods offer the most complete way to perform lag recovery, they are also more complicated and computationally arduous. Aside from costing sheer time to compute, this also opens room for numerical errors (see discussion of the aliasing mixing problem in Section~\ref{sec: RM_theory-aliasing}). For this reason, the less rigorous but more exhaustive family of non-GP methods, which make as few assumptions about the underlying signal properties as possible, are still actively used in conjunction with or in preference over their more statistically precise GP cousins.

Such non-GP methods are valuable in that their lack of loose commitment towards a particular signal model lends them a flexibility and vagueness that absorbs our uncertainty about the AGN signal properties. Such models tend to also be numerically inexpensive, allowing exhaustive searches of their respective parameter spaces forcing their way past potential numerical challenges. Though \litmus is primarily a GP method, we include these here for the sake of completeness and comparison.

Of the extant methods of lag recovery, the Interpolated Cross-Correlation Function \citep[ICCF][]{ICCF-Gaskell_1987}, is conceptually the simplest. The ICCF describes the best fit lag as being the one that maximises the cross correlation, $r$, between the emission and response light curves:
\begin{equation}
    r(\Delta t) =\frac{\braket{y_1 (t), y_2 (t-\Delta t)}}{\sqrt{\braket{y_1(t), y_1(t)} \braket{y_2 (t), y_2 (t)}}},
    \label{eq: cross-correlation}
\end{equation}
where $y_1$ and $y_2$ represent the sets of photometric and spectroscopic amplitudes respectively, and angled brackets indicate an inner product. Because the measurements are not simultaneous, the ICCF reconstructs one or both of the light curves by linearly interpolating between observations. The uncertainty in the recovered lag is estimated from \sqm{bootstrapping}: repeating the lag recovery over multiple realisations generated by randomly sub-sampling observations and re-sampling within their measurement uncertainties.

The ICCF method has been found to agree with more rigorous models like \javelin to within statistical bounds, though with higher reported uncertainties \citep{OZDES-Yu_2019}. Owing to its numerical robustness and low computational overhead, it is still used as a validation tool by both \ozdes \citep{OzDES-Malik_2023, OzDES-Yu_2023, OzDES-Penton_2025} and \sdss \citep{SDSS-Shen_2023} to identify and remove poorly performing for lags recovered by more complex methods. For comparison, \litmus includes a \jax accelerated implementation of the ICCF method in its array of fitting algorithms.

In their final RM data release, \citet{SDSS-Shen_2023} make use of the non-GP method \pyroa \citep{pyroa-ferus_2021}, discussed in more detail in Section~\ref{sec: existing_methods_other}. In this paper we focus on a comparison with the ICCF method, which \ozdes uses as a secondary method as a part of its validation and quality cuts.

\subsection{Other Methods}
\label{sec: existing_methods_other}

Here we give a brief overview of three lag recovery methods that have been widely used outside of \ozdes's RM program: \cream, \mica and \pyroa.

\paragraph{\cream}

The GP-based fitting method \cream \citep{Starkey_2015_CREAM} is similar to \javelin in its modelling of the AGN light curve as a stochastic process and using Bayesian fitting, but differs in two key respects. Firstly, instead of fitting for the parameters of the observed signals it instead fits for physical properties of the AGN (e.g. mass, accretion rate etc.). Secondly it models the light curves by fitting the phase and amplitude of a series of Fourier components and, unlike \javelin, does not include the covariance of the observations in its likelihood function, instead using a $\chi^2$ goodness of fit loss function. A novel feature of \cream is that it can be set to rescale the error bars of measurements from different telescopes as a part of the MCMC fitting in an internal calibration process \citep{starkey_2017_creamerrorscaling}. \cream is used as the primary lag recovery method by some \sdss releases \citep{SDSS-Grier_2017, CIV_Grier_2019}, and is included in the reported results of their final data release \citep{SDSS-Shen_2023}.

\paragraph{\mica} 

Another GP-based fitting method from \citet{Li_2016_MICA} is the code \mica. The main feature of this code is that it constructs the transfer function of Equation~\ref{eq: cc_covar} from a sum of Gaussian profiles with positions and widths as free parameters that are marginalised over in the fitting process. In this way, instead of using a simplified approximate transfer function in the style of \javelin's top-hat, \mica can allow for a more detailed accounting of the BLR's geometry. \mica uses much of the same statistical approach as \javelin, but uses the Metropolis-Hastings algorithm \citep{Metropolis_1953-MetHaste} in place of \emcee.

\paragraph{\pyroa}

Introduced by \citet{pyroa-ferus_2021}, \pyroa adopts a similar strategy to the ICCF but generalises beyond the linear interpolation kernel. It instead uses a \qm{rolling average} to interpolate the light curves, i.e. taking the weighted average with weights following some kernel function, e.g. a Gaussian (multiple options are presented). The width of the kernel is optimised at each point in the reconstructed light curve to maximise the model Bayesian Information Criterion \citep[BIC][]{Schwarz_1978_BIC}, and from these reconstructions the lag is estimated with their correlation function. \pyroa is the main lag recovery method in the final \sdss data release of \citet{SDSS-Shen_2023}, as they find it to be more precise than \PyCCF and more reliable than \javelin. Like \cream, \pyroa includes a utility to re-calibrate measurement uncertainty of observations.

\section{\litmus Methodology}
\label{sec: litmus}

The aliasing problem presents challenges on two fronts. The first is numerical: the posterior distribution is very difficult to properly navigate and map the shape of thanks to its rough and multimodal shape. The second is statistical: how can we tell when a recovery is statistically significant when the aliasing peaks can appear to give meaningfully constrained peaks even in the case of a false positive. Here we present a methodologically consistent solution to to both problems: firstly, to use novel numerical techniques to properly explore the parameter space of a proper Bayesian model, and secondly to use this robust mapping of the posterior distribution to make use of Bayesian model comparison tools to evaluate the significance of a lag recovery in a principled way. In brief: we do the statistics completely, and the numerics properly.

In this section, we introduce \litmus \href{https://github.com/HughMcDougall/litmus/}{\faGithub}, a new lag recovery package, based in \python, making use of modern computational tools like \jax \citep{jax} and its ecosystem of flexible software like the statistical modelling framework \numpyro \citep{numpyro} and the GP modelling package \tinyGP \citep{cite_tinygp}. \litmus's core feature is an algorithm for exploring and integrating the posterior that we call the \hescan, which efficiently explores the parameter space to lags while also finding an estimate of the model evidence such that proper Bayesian null hypothesis testing is possible. \footnote{Note that this method bears some similarity to a special case of the Integrated Nested Laplace Approximation \citep[INLA][]{Rue_2009_INLA,vanniekerk_2022_INLA}.} \litmus also offers a flexible and modular statistical framework that allows it to be easily extended to new generative models, unlike existing tools which hard-code their statistical description of the AGN signals. We then present a new tool for RM that is more precise, mode correct and more robust than anything else in the literature while also being significantly faster and more broadly applicable.

\subsection{The \hescan}
\label{sec: litmus-hessianscan}

The core idea of the \hescan is to side-step the rough geometry of the posterior along the lag-axis (e.g. see the bottom panel in Figure~\ref{fig: aliasing_explained}) by not trying to navigate over this geometry at all. Instead, a grid of test-lags are spaced along the lag-axis, and the shape of the posterior along all non-lag parameters is estimated with the Laplace approximation (approximating a distribution as Gaussian by a second order expansion of its log-density). In contrast to MCMC-like strategies which need to explore each axis of parameter space to marginalise over them, this Laplace approximation approach relies only on optimisation, for which the complexity scales much less severely with dimensionality of the parameter space. As such, the \hescan approach can be applied up to markedly higher model dimensions without incurring an exorbitant computational burden.

\begin{figure}
    \centering
    \includegraphics[width=0.9  \linewidth]{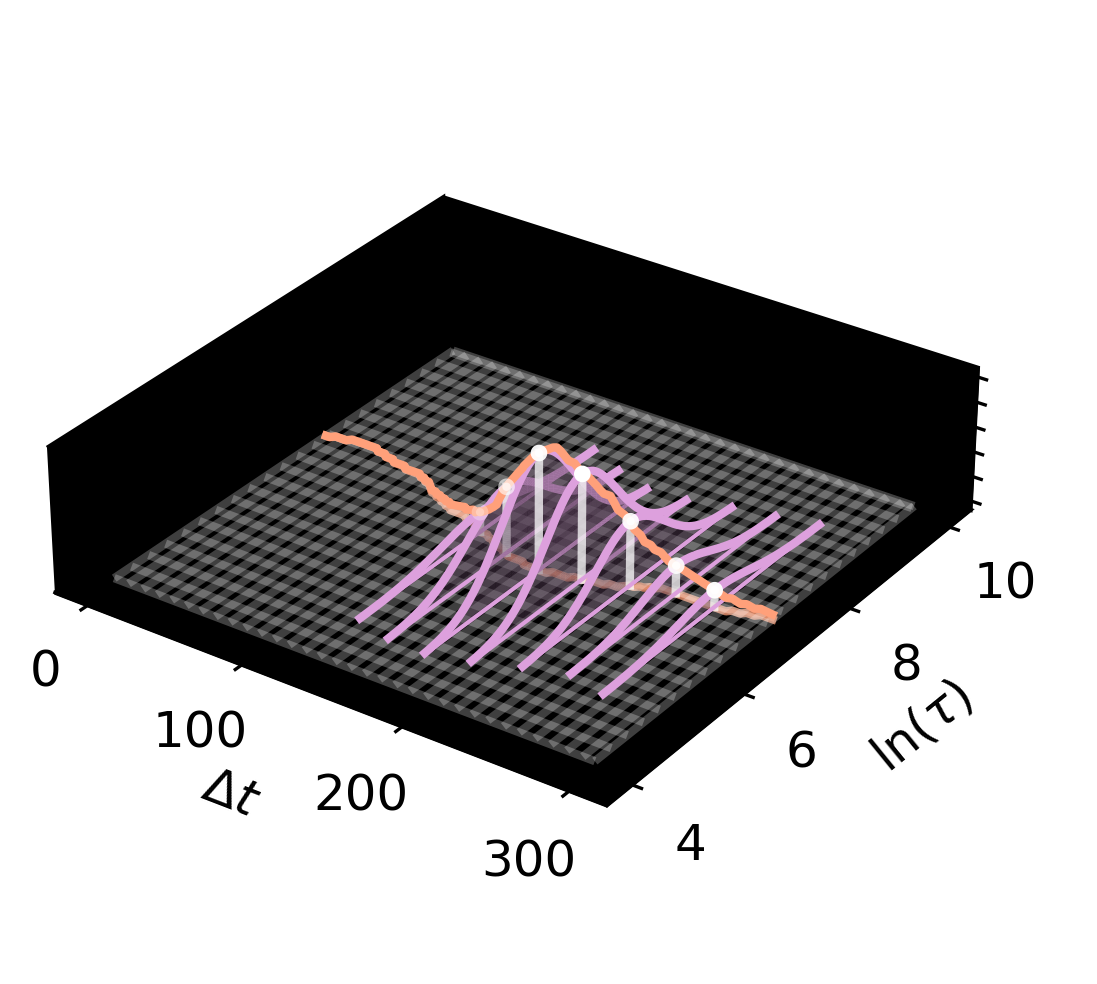}
    \caption{A simplified demonstration of the operating principle behind \litmus's \hescan for a case of only two free parameters (lag and DRW timescale). First, a 1D locus of conditional optima is traced out along the lag axis (orange line), finding the conditional optima at a discrete grid of lags (white points). At these points, the Laplace approximation is applied to divide the posterior up into a series of Gaussian slices (purple, shaded).}
    \label{fig: litmus_slices}
\end{figure}

We begin with a set of $I$ ordered lags, $\{\Delta t_i\}$, distributed, though not necessarily evenly, between the upper and lower ranges of the uniform lag prior:
\begin{equation}
    \Delta t_{\mathrm{Min}} <= \Delta t_0 < \Delta t_1 < \Delta t_2< \ldots <\Delta t_I < \Delta t_{\mathrm{Max}}
    .
\end{equation}
At each of these lags, there is an (here, un-normalised) conditional joint distribution $P_i( \phi)$ for all of the un-fixed non-lag parameters $\phi$, (i.e. the full set of model parameters in Equation~\ref{eq: joint_definition} is $\theta=\{\Delta t\}\cup \phi$):
\begin{equation}
    \mathcal{P}_i( \phi) =  \pi(\phi \vert \Delta t) \mathcal{L}(\phi \vert \Delta t),\\ \; Z_i= \int \mathcal{P}_i(\phi) d\phi
    ,
\end{equation}
where $Z_i$ is the evidence integral of the un-normalised conditional distribution such that $Z_i = \frac{dZ}{\Delta t}\vert_{\Delta t = \Delta t_i}$. The \hescan approximates this distribution and evidence via the Laplace approximation, i.e. supposing that the conditional distribution can be approximated by a multivariate Gaussian distribution:
\begin{equation}
    P_i( \phi) \approx  Q_i(\phi) = \mathcal{N}(\phi \vert \mu , \Sigma)
    ,
    \label{eq: normal_slice}
\end{equation}
where the mean, $\mu$, and covariance matrix, $\Sigma$, of this Gaussian can be estimated from the optimum of the distribution, $\hat{\phi}$, and the Hessian matrix, $H$, of the log density at that point:
\begin{equation}
    \mu =  \hat{\phi}^i, \; \\ \Sigma = -H(\hat{\phi}^i)^{-1}
    .
    \label{eq: Laplace_mu_sigma}
\end{equation}
Here, the Hessian matrix is the curvature of the log-density, i.e.
\begin{equation}
    H_{i,jk}(\phi) = \frac{\partial^2 \logof{\mathcal{P}_i(\phi)}}{ \partial \phi_j \partial \phi_k}
    .
    \label{eq: hessian_definition}
\end{equation}
As \litmus is built in \jax and \numpyro, this Hessian matrix can be evaluated with the need for hand-calculated derivatives. Gaussian distributions have easily evaluable integrals, and so the Laplace approximation also lets us easily calculate $Z_i$ for each slice:
\begin{equation}
    \logof{Z_i} \approx \logof{\mathcal{P}_i(\hat{\phi}^i)} + 
    \frac{1}{2} \logof{\mathrm{det}(\Sigma)}+
    \frac{\mathrm{Dim}(\Sigma)}{2} \logof{ 2\pi}
    .
    \label{eq: Laplace_log_evidence}
\end{equation}
\noindent
To summarise, the \hescan procedure is:
\begin{enumerate}
    \item Generate some set of ordered lags $\{\Delta t_i\}$.
    \item At each lag $\Delta t_i$, optimise all other free parameters $\phi$ to find the conditional optimum $\hat{\phi}^i$.
    \item Calculate the density $\mathcal{P}_i(\hat{\phi}^i)$ and Hessian matrix $H_i(\phi=\hat{\phi})$ of the conditional distribution.
    \item Using these, form a normal distribution approximation for the conditional distribution of the non-lag parameters (the Gaussian slice) with mean and covariance from Equation~\ref{eq: Laplace_mu_sigma} and with normalising evidence from Equation~\ref{eq: Laplace_log_evidence}. This is done with \numpyro's autodiff capabilities.
    \item With the integral at each Gaussian slice, use finite element integration along the $\Delta t$ axis to estimate the full integral / model evidence via Simpson's rule, the Trapezoidal rule, etc.
\end{enumerate} 
A simplified example of this for a case where $\phi = \{\tau\}$ and $I=5$ is shown in Figure~\ref{fig: litmus_slices}. Evenly spaced test lags are wasteful and error-prone, and so \litmus uses a \qm{grid smoothing} algorithm, outlined in Appendix~\ref{app: grid_smoothing}, to preferentially sample good-fit lags. For the optimisation to find $\hat{\phi}^i$ we use \jaxopt's implementation of the Broyden-Fletcher-Goldfarb-Shanno (BFGS) algorithm \citep{Broyden_1970_BFGS01, Fletcher_1970_BFGS02, Goldfarb_1970_BFGS03, Shanno_1970_BFGS04}, using also a pre-conditioning scheme described in Appendix~\ref{app: optimisation}. Assuming that $\hat{\phi}$ is a smooth and continuous function of $\Delta t$, we use the solution $\hat{\phi}^i$ as the starting point for finding $\hat{\phi}^{i+1}$. To find the Hessian matrix, we use \jax's autodiff tool-set on the probability log-density functions provided for a particular model by \numpyro. 

\subsection{Statistical Modelling}
\label{sec: litmus-statsmodelling}

In its statistical modelling, \litmus adopts the same approach as \javelin in describing the AGN fluctuations as a damped random walk and using the Bayesian likelihood in Equation~\ref{eq: GP_likelihood}. For simplicity, we do not include any smoothing of the response function, i.e. we use a transfer function $\psi(t) = \delta(t)$, where $\delta$ is the Dirac delta function. Again following the lead of \javelin, we adopt uniform priors for all variables, except the damping timescale $\tau$ and continuum signal amplitude $\sigma_c$ which are fit on a log-uniform scale owing to their strong correlation and the lack of a prior knowledge of the magnitude of the signal timescale. In this way, \litmus's statistical models are one-to-one with \javelin's under the condition of no smoothing, i.e. $b=0$ in Equation~\ref{eq: tophat_tfer_func}. We can define two simpler models to test the significance of our lag recovery against:
\begin{itemize}
    \item The continuum and response being uncorrelated DRW's, i.e. no lag encoded in the signal, and
    \item The continuum being a DRW, but the response being pure white noise.
\end{itemize} 
The Bayes factors comparing these models allows us to test the significance of a lag recovery. For an example of how these ratios can be interpreted, see Section~\ref{sec: hypothesis_testing}.

\subsubsection{Alternative Statistical Models}
\label{sec: litmus-statsmodelling-altmodels}

A useful feature of \litmus is that, unlike existing tools which hard-code their statistical models, its modular design means that \litmus can also fit alternative statistical descriptions of the signals, e.g. the null hypothesis models used to test the significance of lag recoveries. As an example, \litmus includes an alternative to the normal GP modelling but using an adjustable log-normal prior of the kind that would be recovered from an $R-L$ relationship, i.e. $\pi(\Delta t)\sim \mathcal{N}$. Similar extensions can be made to incorporate the flexible transfer functions of \mica or the physically motivated parameterization and error calibration of \cream, features that \litmus's initial release lacks.

This modularity also allows for more expressive models to probe physical phenomena, for example:
\begin{itemize}
    \item Using a summation of additional GP kernels to examine differences between the signal and the canonical DRW,
    \item Modelling the BLR holiday \citep{Dehghanian_2019}, e.g. with a short-lived signal super-imposed on the GP signal, or
    \item Testing models with time-varying lags to investigate changes in the BLR shape or scale
\end{itemize}
are all implementable in \litmus if they can be expressed with a \numpyro generative model, and can be fit by a suitable choice of its existing fitting methods.

\subsection{Additional Algorithms}
\label{sec: litmus-altmethods}
As well as the \hescan, \litmus includes other lag recovery algorithms for comparison. Firstly, it includes a \jax based high speed implementation of ICCF. Secondly, it includes a variation of the \hescan that uses stochastic variational inference (SVI) in place of the Laplace approximation, and finally it has an interface with the nested sampling \citep{Skilling_2006_NS}, specifically \jaxns \citep{Albert_2020_JAXNS}.

\subsubsection{The \sviscan}

\begin{figure}
    \centering
    \includegraphics[width=0.9\linewidth]{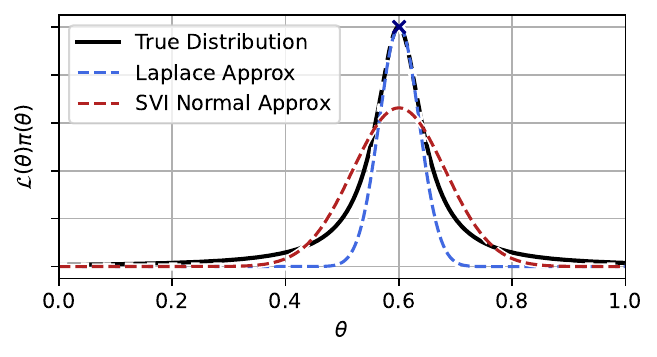}
    \caption{A demonstration of the difference in the Laplace and SVI approximations, both attempting to emulate a Cauchy distribution (black solid line). The Laplace approximation (blue dotted line) creates a Gaussian that matches the curvature at the MAP of the true distribution, and in this case under-estimates the distribution everywhere else. The SVI approximation, here also fitting a Gaussian, instead tries to get as close as possible to the true distribution \sqm{on average}, and so under-estimates in the core region while balancing the impact on the evidence integral with over-estimates in the distribution's tails.}
    \label{fig: SVI_Laplace_example}
\end{figure}

The \hescan relies on Laplace approximation, which can be fragile against highly non-Gaussian posteriors. As an extension \litmus also offers the \sviscan, which has similar working but uses SVI to generate its Gaussian distribution approximation. While the Laplace approximation fits a Gaussian with a point-estimate based on the Hessian of the log-posterior at its maximum, SVI seeks to fit a Gaussian that is \sqm{most similar on average} (see Figure~\ref{fig: SVI_Laplace_example} for a visual demonstration). This is codified by the Kullback–Leibler divergence (KL divergence), or \qm{average log difference} between the Gaussian \sqm{surrogate distribution} in Equation~\ref{eq: normal_slice} and the true posterior density:
\begin{equation}
    \mathrm{KL}_{P_i (\phi) Q_i(\phi)} 
    = \mathbb{E} \left[ \logof{\frac{P_i(\phi)}{Q_i(\phi)}} \right]_{Q_i(\phi)}
    ,
    \label{eq: KL_definition}
\end{equation}
where $P_i(\phi)$ is again the un-normalised joint distribution conditioned at test lag $\Delta t^i$. The slice evidence $Z_i$ is not known a priori, but can be factored out of the expression, such that the minimum value of $\mathrm{KL}$ can be found to within an additive constant:

\begin{equation*}
    \mathrm{KL}_{P_i (\phi) \rightarrow Q_i(\phi)} 
    = \mathbb{E} \left[ \logof{\frac{\mathcal{P}_i(\phi)}{Q_i(\phi)}} \right]_{Q_i(\phi)} - \logof{Z_i}
\end{equation*}

It can be shown that the KL divergence is always positive, which means that $\logof{Z_i}$ is always greater than or equal to the expectation value in Equation~\ref{eq: KL_definition}, being equal at a perfect match between the surrogate distribution and the true posterior (i.e. when $\mathrm{KL}$=0). As such, this is called the \qm{Evidence Lower Bound} (ELBO) of the fit, and can be approximated by evaluating the log-difference at samples drawn from $Q_i(\phi)$:
\begin{equation}
    \logof{Z_i} \ge 
     \mathrm{ELBO}_{P_i (\phi) \rightarrow Q_i(\phi) }
     \approx \frac{1}{M} \sum_{m=1}^{M}{\logof{\frac{\mathcal{P}_i(\phi_m)}{Q_i(\phi_m)}}}, \; \phi_m \sim Q_i(\phi_m)
     \label{eq: ELBO_summation}
\end{equation}

By maximising the $\mathrm{ELBO}$ we find both the highest (best) estimate of the evidence, and also the surrogate distribution that is the closest fit to the posterior. This is done by assuming some parametric ansatz form for the surrogate distribution and optimising these parameters. In our case we use the normal distribution in Equation~\ref{eq: normal_slice} with parameters being the mean and covariance. In actuality the symmetry of the covariance matrix means that we optimise the elements of its Cholesky decomposition, $L$, where $\Sigma=LL^T$. 

The exact value of the $\mathrm{ELBO}$ is not easily evaluable, and is instead estimated from the summation in Equation~\ref{eq: ELBO_summation}. This is a necessarily stochastic measure as it relies on a finite set of random samples. As such maximising the $\mathrm{ELBO}$ makes use of stochastic optimisation. In \litmus's case, use \numpyro's ready-built SVI tool set and the stochastic optimiser \sqm{adam} \citep{kingma_2017_ADAM}. Aside from using SVI to estimate the Gaussian slices and the $\mathrm{ELBO}$ to estimate the slice evidence, the \sviscan operates identically to the \hescan in Section~\ref{sec: litmus-hessianscan}. 

\subsubsection{Nested Sampling}

Nested sampling \citep{Skilling_2006_NS} operates by taking a uniformly distributed ensemble of \sqm{live points} along with a statistical estimate of the volume of parameter space that they subtend. Its iterations slowly shrink and split the volume subtended by this ensemble, at each step keeping track of changes to this volume. The result is an ordered series of Lebesgue integral elements describing the nominal posterior density and volume of a series of contours, the summation of which estimates the evidence integral. 

In \litmus we include nested sampling by way of \jaxns, a \jax-accelerated implementation that combines the Gaussian shell approach of \texttt{MultiNEST}\xspace\citep{Feroz_2009-Multinest} with the Slice sampling of \texttt{PolyChord}\xspace \citep{Handley_2015-Polychord}. Unlike the Laplace and SVI quadrature methods, no assumptions of Gaussanity are made, and so this approach is unconditionally convergent in the limit of large sampling. For the specifics of Nested Sampling, including its convergence diagnostics and uncertainty estimates, we direct the reader to \citet{Ashton_2022} for a more detailed review.

\subsection{Hypothesis Testing}
\label{sec: hypothesis_testing}

\begin{figure*}
    \centering
    \includegraphics[width=0.45\linewidth]{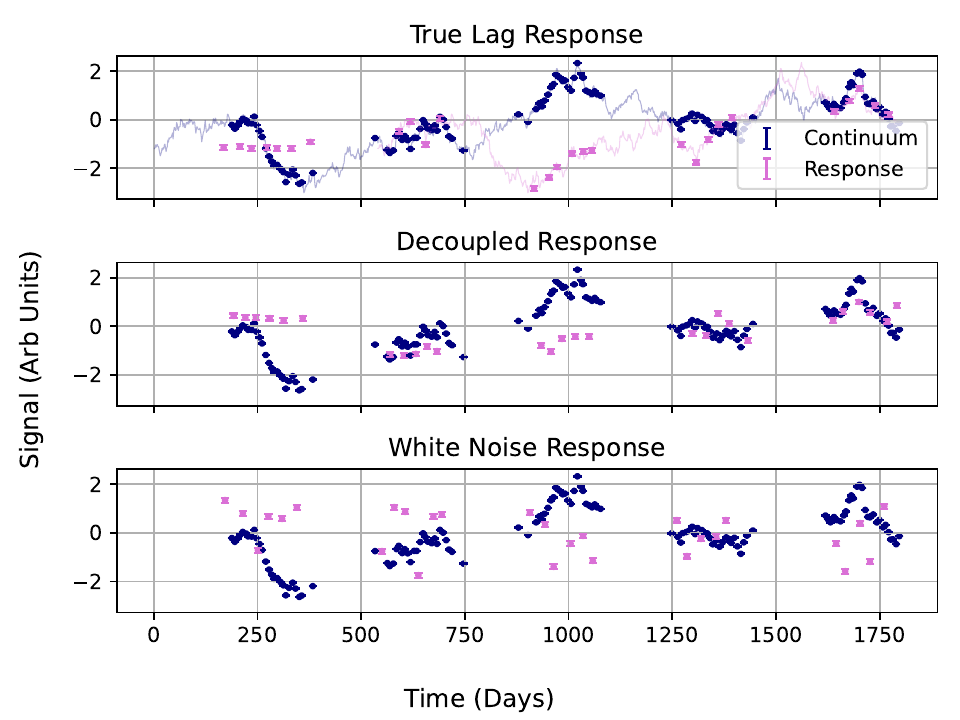}
    \includegraphics[width=0.45\linewidth]{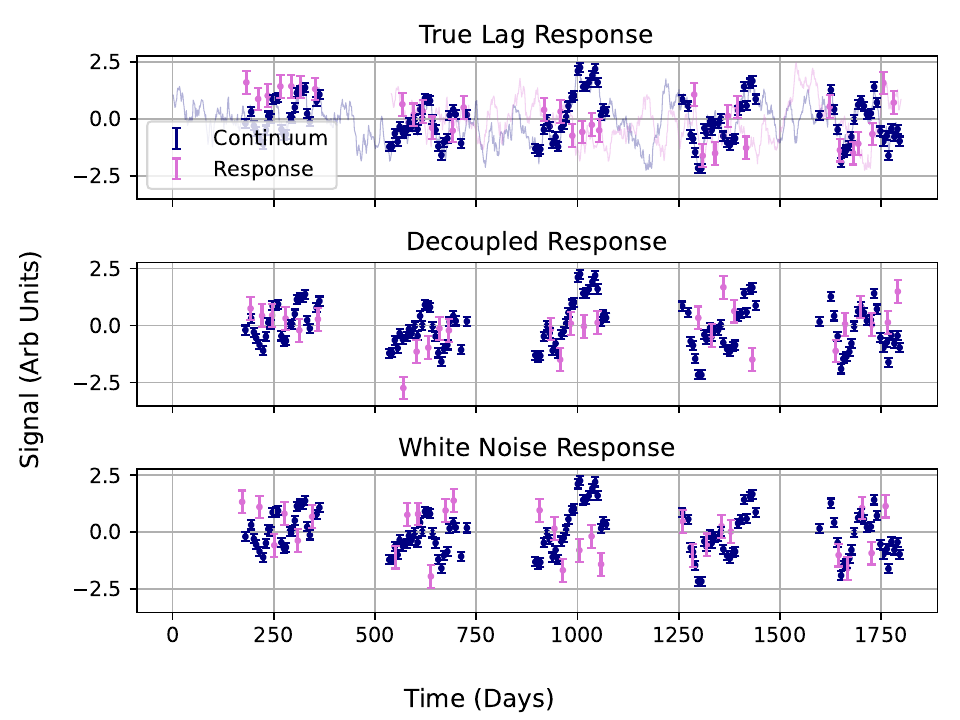} \\ 

    \includegraphics[width=0.95\linewidth]{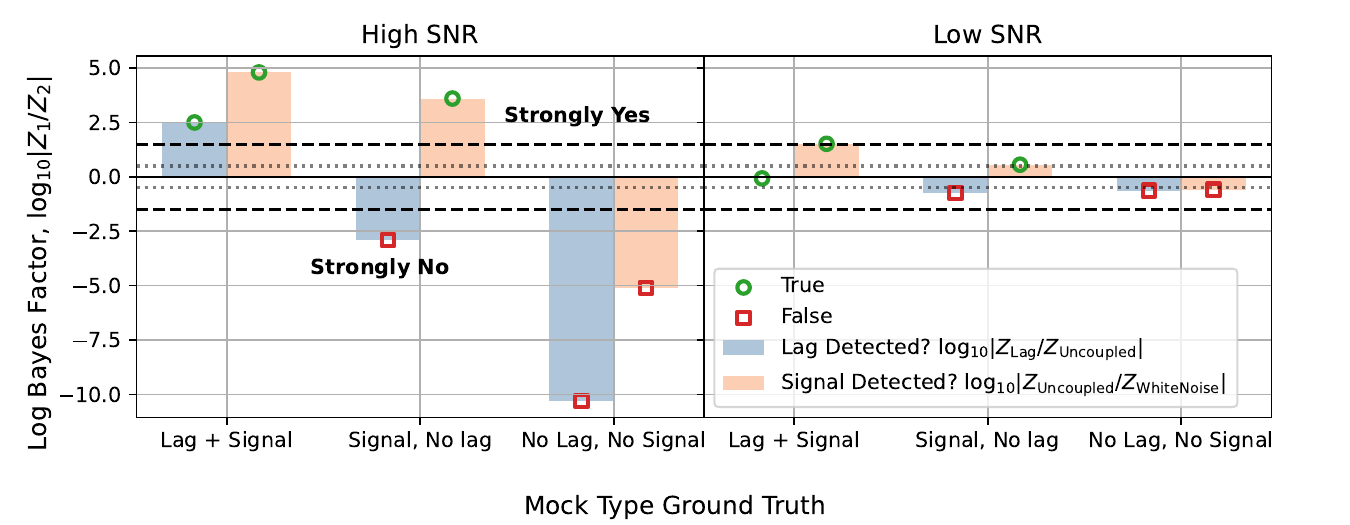}

    \caption{Mocks and evidence ratios for the demonstrative mock signals in the body of the text. The top panels shows the mock light curves for continuum (blue) and response (pink) signals. The left and right columns correspond to high and low SNR, while the rows from top to bottom show the mocks for a coupled response at a lag of $\Delta t = 540 \dayu$, a decoupled response and a pure white noise response. The bottom panel shows the differences in log-Bayes factors for each of the hypotheses in Table\ref{tab: null_test}, with orange bars being how strongly we can see structure in the response signals and blue bars being how well we can confirm the existence of a lag from this structure, with dotted lines indicating different significance levels in favour of accepting or rejecting these hypotheses. Red and green markers dots indicate the ground truth for each question: green circles mean true, red squares mean false.}
    \label{fig: null_test}
\end{figure*}

\litmus's biggest improvement over existing methods is its ability to not only constrain the best fit lag, but determine the significance of there being a lag at all. We can demonstrate this feature on mock data, using models from Section~\ref{sec: litmus-statsmodelling} and Bayesian evidence using the algorithms in Section~\ref{sec: litmus-hessianscan} and Section~\ref{sec: litmus-altmethods}. As an example we can use the prickly $\Delta t = 540 \dayu$ case, shown in Figure~\ref{fig: null_test}, which falls into one of the aliasing gaps and so is maximally ambiguous. At this lag we generate six mock cases, shown in the top panel, broken into high SNR and low SNR examples:
\begin{itemize}
    \item A continuum and response that actually encodes a lag.
    \item That same continuum, but with a response from a different mock such that it is still a DRW with the same timescale, but encoding no meaningful lag response.
    \item The same continuum again but with a response signal that is pure white noise, encoding no signal structure whatsoever.
\end{itemize}
In this way we can use Bayes factors (ratios of the models' Bayesian evidences) to test null hypotheses via model comparison. Here we formalise this process as asking two questions: is there a signal in the response, and if so does this signal demonstrate a significant lag? We consider a "strong" result to be a Bayes factor of $100\times$ or more, and a moderate result to be a factor of $10\times$ or more.  Using the \hescan we find that we successfully identify the true lag where it exists, and our evidence calculations allow us to identify and exclude the spurious recoveries for the two false mocks at high SNR. 

Figure~\ref{fig: null_test} contains a demonstration of \litmus's ability to use Bayes factors to distinguish if a signal is lag-carrying and/or contains structure. We simulate mock light curves emulating the cadence and uncertainty of \ozdes in two cases: mocks with a low measurement uncertainty and easily observable slow fluctuations with a variational timescale signal of $\tau=200 \dayu$, and mocks with a higher measurement uncertainty and a more rapid variational timescale of $\tau=50 \dayu$ that is difficult to distinguish from white noise with such a coarse observational cadence. For each case we simulate a single continuum but three different response signals: a simulated reverberation response with a lag of $\Delta t = 540 \dayu$, a response that follows the same DRW but is decoupled such that there is no observable lag (equivalent to $\Delta t \rightarrow \infty$) and a response that is drawn purely from white noise (equivalent to a variational timescale of $\tau=0$). Using the \hescan's evidence integrals, \litmus can successfully recover the truth for all tests at strong significance (all true positives and true negatives) in the high SNR case, and returns no false positives even when the constraints are weak.

\begin{table*}
    \footnotesize
    \centering
    \begin{tabular}{|l|c|c|c||c|c|c|}
        \hline \hline
        \multirow{3}{*}[-4pt]{\thead{Mock Type}} & \multicolumn{6}{c|}{
        $\bm{\log_{10} \lvert Z/Z_{\mathrm{White Noise}}\rvert}$
        } \\ \cline{2-7}

        & \multicolumn{3}{c}{Low SNR Mocks} & \multicolumn{3}{c|}{High SNR Mocks} \\ \cline{2-7}
        
        &\makecell{Lag \\ + Structure} & \makecell{Structure, \\ No Lag}& \makecell{No Lag \\ or Structure}
        &\makecell{Lag \\ + Structure} & \makecell{Structure, \\ No Lag}& \makecell{No Lag \\ or Structure}
        \\ \hline

        Lag  Response &7.3   &  4.8 & 0.0  &1.5   &  1.5 & 0.0 \\\hline
        Uncoupled Response &0.7   &  3.6 & 0.0 &0.2   &  0.6 & 0.0 \\\hline
        White Noise Response 	 &-15.4 & -5.1 & 0.0 &1.2 & 0.6 & 0.0 \\\hline\hline
    \end{tabular}
    \caption{Bayes factors (log scale) of model evidences when the different mocks in Figure~\ref{fig: null_test} are fit with a model that encodes a lag response or that encodes an uncoupled but still structured response signal as compared to a model in which the response is unstructured noise. The bottom panel of Figure~\ref{fig: null_test} shows the Bayes factors from these evidences that are used to test different hypotheses / compare the relative strength of the different models.}
    \label{tab: null_test}
\end{table*}

\section{Validation of Results}
\label{sec: validation}

The example in Section~\ref{sec: hypothesis_testing} demonstrates \litmus's success on a small number of mocks, but it is necessary to also show that it performs reliably at scale. In this section, we test and compare \litmus's performance against existing methods on three sets of mocks:
\begin{enumerate}
    \item \label{mock_uniform}A set of $440$ mock AGN with lags distributed uniformly in $\Delta t \in [0,1000] \; \dayu$ and timescales drawn log-uniformly in $100-1000 \; \dayu$.
    \item \label{mock_normal}A similar sample of $490$ mocks with lags drawn from a log-normal distribution, $\Delta t \sim \mathcal{N}(2,0.4)$, roughly emulating the spread of the \ozdes \mgii sample.
    \item \label{mock_false}The same set of $490$ mocks in set~\ref{mock_normal} but with the response curves randomly regenerated to create lag-free mocks, similar to the decoupled responses in Fig~\ref{fig: null_test}.
\end{enumerate}
Mock set~\ref{mock_uniform} allows us to interrogate \litmus's performance in lag recovery and its resistance to aliasing, particularly in contrast to \javelin's AEIS method, while mock sets~\ref{mock_normal} and \ref{mock_false} allow us to examine how the Bayes factor allows us to screen for false positive lags. The decoupled mocks are of physical interest as they correspond to the case of the lag being longer upper limit detectable in a particular survey, as can happen for longer observer-frame \civ lags \citep{OzDES-Penton_2025}. Mocks are all generated with continuum and response amplitudes of unity and zero mean, with weekly cadence and $\approx 1\%$ uncertainty for the continuum observations and monthly cadence with $\approx10\%$ uncertainty for the response light curves. 

For each mock in each set, we recover the lag with \litmus's \hescan, \sviscan and Nested Sampling fitting methods, as well as its in-built ICCF and \javelin-like AEIS fitting methods.\footnote{The AEIS method does not map exactly to a true \javelin fit as no smoothing is applied from a transfer function and fitting is performed in the unconstrained domain (See Appendix~\ref{app: litmus-statsmodelling-constrained}).} The fitting parameters for each method are listed in Appendix~\ref{app: speed_and_tuning}. For a fair comparison we only include mocks for which all five fitting methods are have fully converged. For the Nested Sampling method we use $10^4$ live points, enough to ensure good performance, so that this well established algorithm can act as a benchmark to test the other four fitting methods against. In this benchmarking, use the rough metric of a lag being correct or incorrect if it agrees / differs from the true lag by a threshold of $30 \; \dayu$.

\begin{figure*}
    \centering
    \includegraphics[width=0.9\linewidth]{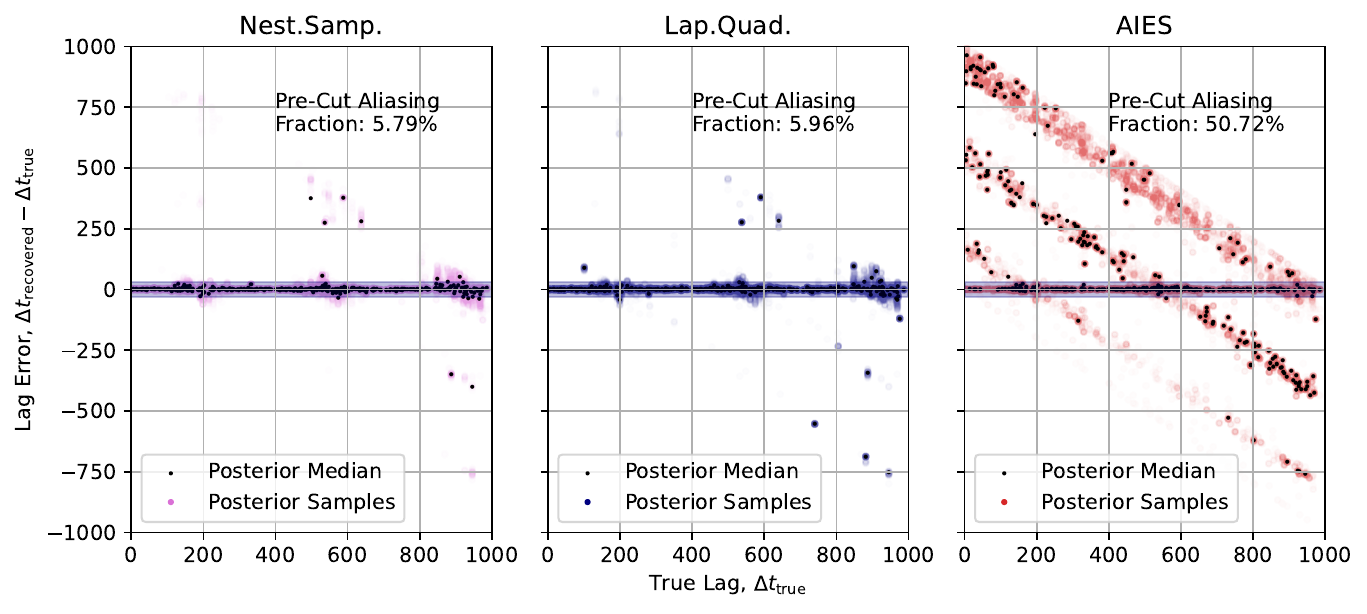}
    \caption{A comparison of the posterior distributions for the lag error, i.e. the difference between true and recovered lag, comparing some of \litmus's aliasing-friendly methods, namely Nested Sampling (left panel) and the \hescan (middle panel), to the \javelin-like AEIS (right panel). These plots are for mock sample~\ref{mock_uniform} which has $440$ mocks with true lags distributed uniformly over the range $\Delta t \in [0,1000] \;\dayu$. The aliasing fraction is the fraction of samples / posterior density that sits more than $30 \; \dayu$ from the true value for these mocks. The \hescan and Nested Sampling results adhere extremely closely to the true lags save for a single errant false positive, and the similarity between the two validates the \hescan's recovery of the true posterior shape. Conversely, more than half the AEIS samples are incorrect, and the posterior median (black dots) is often far from the true value.}
    \label{fig: aliasing_fraction}
\end{figure*}

As can be seen in Figure~\ref{fig: aliasing_fraction}, \litmus's aliasing friendly fitting methods produce posteriors that map closely to the true mock lags save a few outliers that can be mostly screened with significance tests (see Table~\ref{tab: Benchmark_results}). We can also see that the \hescan's assumption of Gaussianity does not significantly disrupt this result as compared to Nested Sampling. Conversely, the results from the \javelin-like AIES are heavily obscured by the impacts of aliasing, with more than half the posterior samples sitting in the seasonal striations of the aliasing bands. Even before screening spurious lags with Bayes factor tests, \litmus's various fitters are significantly better than the AEIS at identifying correct lags.

\begin{figure}
    \centering
    \includegraphics[width=0.9\linewidth]{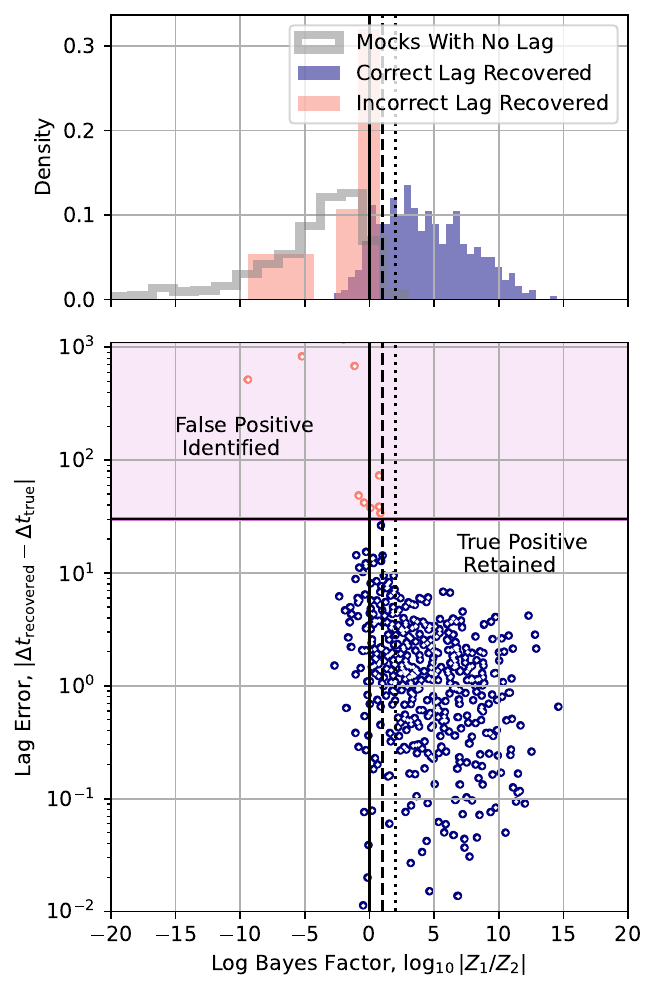}
    \caption{A demonstration of how the Bayes factor acts as a measure of lag reliability. The top panel shows histograms of the Bayes factor evidence ratios for the decoupled mocks with no lag (grey), mocks with a lag that was successfully recovered (navy) and mocks that had an underlying lag but for which the posterior median of the recovery was more than $30 \; \dayu$ from the ground truth. The bottom panel shows how the error in the lag (here, the deviation between ground truth and posterior median) decreases for strong Bayes factors. As the evidence ratio for the lag and decoupled models lowers, the error in the median recovered lag rapidly increases, and above some reasonable threshold (e.g. $Z_2/Z_1>10^2$, the results become significantly more reliable. The solid, dashed and dotted lines represent evidence ratio thresholds of $1{:}1$, $1{:}10$ and $1{:}100$ in favour of a lag. The correct and incorrect lags are for the $490$ realistic mocks in mock set~\ref{mock_normal}, while the mocks with no lag are from the decoupled mock set~\ref{mock_false}.}
    \label{fig: SNR_performance}
\end{figure}

In Figure~\ref{fig: SNR_performance}, bottom panel we can see how the Bayes factor acts as a measure of the reliability of lag recoveries. As the Bayes factor between the models of lag-bearing and decoupled response becomes smaller, the error in the recovered lag (measured from the posterior median) rapidly increases. Figure~\ref{fig: SNR_performance} also shows in its top panel how the Bayes factor helps to separate signals that do and do not encode lags within our prior range, with the decoupled population (mock set~\ref{mock_false}) sitting almost entirely at Bayes factors of $Z_2/Z_1<1$. This shows how the Bayes factor offers a direct and easily implementable means screening low confidence lag recoveries, and how this quality cuts can be tuned to a nominal level of reliability.

We compare all of \litmus's fitting methods, in concert with a quality cut of enforcing $Z_2/Z_1\ge10$, to the existing methods of lag recovery (\javelin-like AEIS and \PyCCF-like ICCF) in Table~\ref{tab: Benchmark_results}. To roughly emulate the quality cuts of the \mgii \ozdes sample, we employ the first $2$ of \citet{OzDES-Yu_2023}'s quality cuts on the AEIS results, namely that the width of the AEIS posterior, as measured between the 16\textsuperscript{th} and 84\textsuperscript{th} percentiles, is less than $110 \; \dayu$, and that the AEIS and ICCF methods agree to within $2\sigma$, where $\sigma$ is the standard deviation of the AEIS posterior. For the ICCF results we retain only results that agree with the AEIS to within $100 \; \dayu$, and for which the standard deviation of the bootstrapped lag recoveries are also less than $100 \; \dayu$. 

Prior to applying cuts, \litmus's alias-friendly methods identify the correct lag in both the uniform and log-normal samples with a False Positive Rate (FPR) of $<5\%$, while the ICCF and AEIS methods recover the incorrect lag at an FPR of $\approx40\%$ or higher in these mock sets. After discarding sources with a Bayes factor $Z_2/Z_1<10$, the \litmus results improve even further, and importantly retain the vast majority of true recoveries. By contrast the AEIS and ICCF methods, while still seeing a decrease in FPR, do so at the cost of discarding much of the sample. Similarly in the entirely decoupled mock set~\ref{mock_false}, which contains only false positives, the Bayes factor test removes $\approx98-99\%$ of these sources while the ICCF and AEIS cuts retain almost ten times more. In short, \litmus's methods identify the correct lag more often and performs much better at discarding incorrect lags. 

Note that these result should not be considered indicative of the actual FPR in the published \ozdes \hbeta, \mgii and \civ lags, as these mocks samples to not fully emulate the physical \ozdes sample, nor do they apply all of \ozdes's stringent selection criteria. With that being said, the improved performance of \litmus's methods in fitting lags over ICCF and AEIS are clear.

\begin{table}[]
    \footnotesize
    \centering
    \begin{tabular}{|l|c|c|c|}
    \hline\hline
         &  \thead{\makecell{FPR Before \\ Cuts}} &  \thead{\makecell{FPR After \\ Cuts}} &  \thead{\makecell{Retained \\ Sources}} \\\hline

        \multicolumn{4}{|c|}{Mock Set~\ref{mock_uniform} - Uniform Lags} \\ \hline

         \hescan &  $4.32\%$ & $0.57\%$ & $80.23\%$ \\\hline
         \sviscan &  $4.09\%$ & $0.59\%$ & $77.27\%$ \\\hline
         Nest. Samp. &  $3.18\%$ & $0.00\%$ & $52.05\%$ \\\hline
         \javelin-Like AIES &  $41.59\%$ & $12.90\%$ & $14.09\%$ \\\hline
         ICCF &  $43.41\%$ & $30.34\%$ & $20.23\%$ \\\hline\hline

         \multicolumn{4}{|c|}{Mock Set~\ref{mock_normal} - Log-Normal Lags} \\ \hline

         \hescan &  $4.29\%$ & $1.30\%$ & $94.08\%$ \\\hline
         \sviscan &  $2.45\%$ & $1.07\%$ & $95.10\%$ \\\hline
         Nest. Samp. &  $2.24\%$ & $0.00\%$ & $79.39\%$ \\\hline
         \javelin-Like AIES &  $66.73\%$ & $61.54\%$ & $79.39\%$ \\\hline
         ICCF & $39.39\%$ & $18.75\%$ & $9.80\%$ \\\hline\hline
         
         \multicolumn{4}{|c|}{Mock Set~\ref{mock_false} - Decoupled Response} \\ \hline

         \hescan & - & - & $0.98\%$ \\\hline
         \sviscan & - & - & $0.78\%$ \\\hline
         Nest. Samp. & - & - & $1.37\%$ \\\hline
         \javelin-Like AIES & - & - & $13.28\%$ \\\hline
         ICCF & - & - & $8.01\%$\\\hline\hline 
    \end{tabular}
    \caption{Summary of the performance of \litmus's three fitting methods, the \hescan, \sviscan and Nested Sampling approaches, as compared the AEIS fitting method used by \javelin and the ICCF method use by \PyCCF, as tested on the three sets of mock light-curves (uniform in lag, log-normal in lag to emulate the \ozdes $\mathrm{MgII}$ sample, and mocks with the response light curve decoupled such that no true lag is present). Listed are the fraction of false positive lags before and after quality cuts (a lag here being considered incorrect if it differs by more than 30 days from the ground truth), as well as the total number of retained sources after cuts. In general, the \litmus fitting methods perform significantly better at identifying a true lag where it exists, with a pre-cut $\mathrm{FPR} < 5\%$ in all cases, reducing by a factor of a few when removing sources with a lag recovery evidence ratio $Z_2/Z_1<10$. Overall, \litmus yields significantly more and and significantly more accurate lags, while also retaining $10-20 \times$ fewer spurious lags from the decoupled sample.}
    \label{tab: Benchmark_results}
\end{table}

\section{Discussion \& Future Work}
\label{sec: discussion}

\litmus's improved handling of the aliasing problems of multi-year RM surveys means that, by being applied to existing data, it stands to improve the statistical power of our RM results without the need for new observations. \citet{OzDES-McDougall_2025} found that the collective constraining power of the entire literature of RM measurements is sufficient to constrain the high redshift $R-L$ relationships such that the statistical uncertainty is subdominant compared to the inherent population noise. However, it is not confirmed whether a single $R-L$ relationship is sufficient to explain all AGN behaviour over all time. Current constraints are not enough to test the possibility of time-varying $R-L$ relationships, and the existing sample is restricted to only a narrow window of redshift-luminosity space. Through \litmus's hypothesis testing we can not only successfully recover more lags, but also expand their breadth in parameter space, and so will be equipped to begin using RM as a probe of more expressive physical models. 

One such question is \citet{OzDES-McDougall_2025}'s finding that the lags associated with the \mgii line were systematically and significantly larger than those of \hbeta, indicating that the \mgii emission region of the BLR may be exterior rather than cospatial with that of \hbeta, contrary to existing observations \citep[e.g.][]{MgII_Shen_2019}. This result was significant at barely $2\sigma$, meaning even a small increase in the number of \mgii lags would could confirm or falsify this finding. \ozdes and \sdss have both completed their final RM data releases, meaning a sample of nearly $2000$ AGN light curves are 
available for \litmus to be applied to. 

\litmus is a general purpose single-lag RM tool, and so can be applied beyond this BLR RM domain. The same GP statistical model can be applied to lensed quasar time delay without much alteration \citep[e.g. see the GP based approaches in][]{Ding_2021_TDLMC2}, as well as any other single-lag estimation applications. Because \litmus's statistical models can be extended or swapped out, it can also be aimed at answering some of the stickier questions of BLR RM, such as means of accounting for the BLR holiday \citep{Dehghanian_2019} or descriptions of outlier measurements.

\section*{Acknowledgements}
Parts of this research were conducted by the Australian Research Council Centre of Excellence for Gravitational Wave Discovery (OzGrav), through project number CE230100016. HMG \& BJSP acknowledge the support of the DECRA fellowship DE210101639, funded by the Australian Government through the Australian Research Council.

The authors would like to thank Louis Desdoigts for his experience and advice in writing \litmus, and Rahma Alfarsy of the DESI collaboration for aiding in testing its initial stages. We thank also Madeline L. Cross-Parkin for her assistance in formatting the plots in this paper and \litmus's visual style. All software was made using \texttt{python} \citep{VanRossum_2009_python} and with the aid of \texttt{numpy} \citep{harris_2020_numpy}. Plots and figures were generated with the aid of \texttt{matplotlib} \citep{Hunter_2007_matplotlib} and \texttt{chainconsumer} \citep{Hinton_2016_chainconsumer}.

We acknowledge and pay respect to the traditional owners of the land on which the University of Queensland is situated, upon whose unceded, sovereign, ancestral lands we work. We pay respects to their Ancestors and descendants, who continue cultural and spiritual connections to Country. 

We acknowledge the traditional custodians of the Macquarie University land, the Wallumattagal clan of the Dharug nation, whose cultures and customs have nurtured and continue to nurture this land since the Dreamtime.  We pay our respects to Elders past and present.

\section{Data availability statement}
For data availability, we can now link to a \href{https://zenodo.org/records/18253331?token=eyJhbGciOiJIUzUxMiJ9.eyJpZCI6IjZlNmJjMWQ1LWU5Yz ItNDc1ZC1hNjJlLTQwOGUzN2I1ZWU2MiIsImRhdGEiOnt9LCJyYW5kb20iOiJkNGQ4YmNiNDdiNDFhYTcyMWJkZDYzODcwMTIxOTgwZCJ9.N7TU VJUIEe7d72x-6XiRwuZdFWXnDhGj78Cy8DkF_xYEJlgF8zSjHnOI2jq-H0JR2KrUoristry1jOqAm76p_A}{zenodo archive} of the mocks/posterior results and, though the software is linked elsewhere, we also include a link to the \href{https://hughmcdougall.github.io/litmus/}{github} here as well.

%%%%%%%%%%%%%%%%%%%%%%%%%%%%%%%%%%%%%%%%%%%%%%%%%%
%\printendnotes
%\printbibliography
%%%%%%%%%%%%%%%%% APPENDICES %%%%%%%%%%%%%%%%%%%%%
\appendix

\section{The Constrained \& Unconstrained Domain}
\label{app: litmus-statsmodelling-constrained}

Many Bayesian fitting algorithms encounter difficulties with hard boundaries on model priors, and perform best when parameters are \qm{unconstrained}, i.e. with a domain of $\theta \in \mathbb{R},\; -\infty<\theta<\infty$. Such hard boundaries have infinite gradients in log-probability and other pathologies that can lead to artefacts or non-convergence with some fitting methods. This is a particular concern in RM, where convention is to use bounded uniform priors on all parameters. \numpyro obviates these difficulties with a statistical slight of hand. Prior to fitting, all parameters with constrained priors are shifted with a change of variables into a new coordinate space in which they are unconstrained, e.g. for a Uniform distribution on $\theta \sim U_{a,b}(\theta)$ there's a mapping $f: (a,b) \rightarrow \mathbb{R}, \; \theta'=f(\theta)$.

KL divergences and evidences are preserved across this transformation, which changes the posterior density like:

\begin{equation}
    P(\theta) = P(\theta)\times{J(\theta)},
\end{equation}

\noindent where $J(\theta)$ is the Jacobian of the transformation $\theta \rightarrow \theta'$. \litmus's \hescan and \sviscan fitting methods (see Section~\ref{sec: litmus-altmethods}) are all performed in this coordinate system by default, i.e. they assume that the posterior is approximately Gaussian in this unconstrained domain. Strictly speaking all densities should be written in terms of $\phi'$ and $\theta'$ in the description of these algorithms, but we exclude this for simplicity of reading.

\section{The Grid Smoothing Algorithm}
\label{app: grid_smoothing}

\begin{figure}
    \centering
    \includegraphics[width=1.0\linewidth]{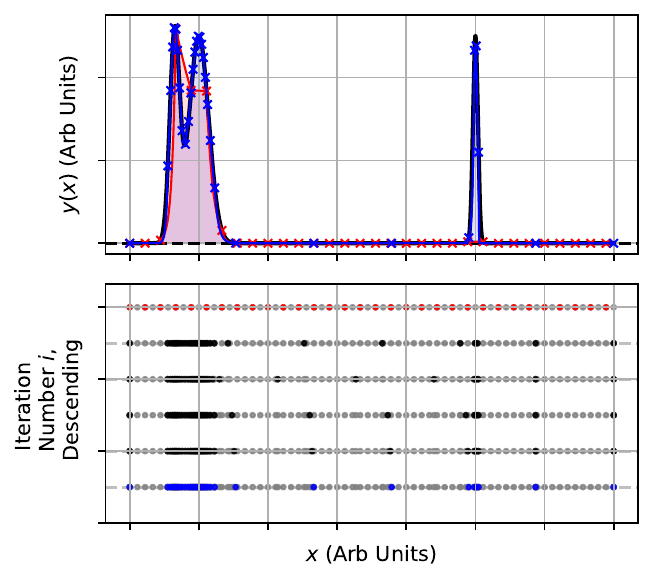}
    \caption{An exaggerated demonstration of the grid smoothing algorithm for a simple multimodal function using $\alpha = 0.8$ up to $j=5$ iterations with $32$ points. The top panel shows the true distribution (black) with its estimate from the first evenly spaced grid (red) and the final smoothed grid (blue). The bottom panel shows how the spacing of the grid updates over each iteration, progressing from top to bottom, with the first and last iterations coloured for emphasis, and gray dots representing samples from previous iterations. The initial spacing is so coarse that it misses much of the detail of the left mode, and cuts off the right-mode entirely. By final iteration, the estimate of the mode is significantly more accurate. }
    \label{fig: grid_smoothing}
\end{figure}

The \hescan and \sviscan algorithms divide the lag domain into an uneven grid of Gaussian slices. If these are spaced too sparsely we can miss important information about the posterior peaks, particular in the narrowly constrained modes of high SNR cases, and if they are packed too densely we can lose information about the long tails of the distribution, which can impact the accuracy of the evidence estimate (see Figure~B-\ref{fig: grid_smoothing}). In the the idealised case, the two extremes are to have the points evenly spaced across the entire domain ($\delta\Delta t$ is constant) and having them distributed with spacing inversely proportional to the posterior density ($\delta \Delta t \propto P(\Delta t)$). 

In \litmus we describe a sliding scale between these with a \qm{grid bunching} parameter, $\alpha$, where $\alpha=0$ is an evenly spaced grid and $\alpha=1$ is bunched proportional to the posterior. Values in between describe spacings whose cumulative distribution function (CDF) is a weighted average of these two extremes. First, we find some estimate of the best-fit parameters for the entire model, $\hat{\theta}$, and from this point forward fix all non-lag parameters at these values, $\hat{\phi}$, to produce a conditional distribution $\propto P(\Delta t | \hat{\phi})$. Estimating the entire marginal distribution $P(\Delta t)$ would be equivalent to solving the entire distribution, and so we form the grid based only on estimates of this much simpler conditional distribution.

For brevity, define 

\begin{equation}
x = \frac{\Delta t - \Delta t_{\mathrm{Min}}}{\Delta t_{\mathrm{Max}} - \Delta t_{\mathrm{Min}}}
\end{equation} 
\noindent and function $y(x) = \mathcal{P}(\Delta t(x)| \hat{\phi})$. The grid smoothing algorithm is as follows:

\begin{enumerate}
    \item Start with a grid of $I$ evenly spaced lags along $x\in[0,1]$ and calculate $y(x)$ for each of these. These obey a cumulative distribution function (CDF) $Y^0(t) = x$, the CDF for initial iteration $j=0$. 
    \item Linearly interpolate between all points and from this estimate the CDF of y: 
    $Y=\int_0^x y(x') dx' / \int_0^1 y(x') dx'$.
    \item Define a \sqm{half-step CDF}, $Y^{j+\frac{1}{2}}(x)$, that is the weighted average of this and a uniform distribution:
    $Y^{j+1}(x) = \alpha Y^j(x) + (1-\alpha) x$
    \item Draw a new set of $I$ points whose spacings obey $Y^{j+1}(x)$, append these to the set of test lags.
    \item Repeat steps (ii)$\mathrm{\rightarrow}$(iv) until converged, usually only a few steps.
\end{enumerate}
Once converged (in Figure~\ref{app: grid_smoothing} this is after $5$ iterations), we construct a grid of $I$ ordered $x$ values from the final $Y(x)$ and convert to the desired set of lags ${\Delta t^i}$ with $\Delta t^i = \Delta t_{\mathrm{Min}} + x_i^{J-1} (\Delta t_{\mathrm{Max}} - \Delta t_{\mathrm{Min}})$.

We find that $\alpha\in [0.5,0.8]$ tends to give the best results over a wide range of lag posteriors. 

\section{Optimisation, Preconditioning \& Convergence Tests}
\label{app: optimisation}
The \hescan algorithm's speed comes from its shifting of the integral problem into an optimisation problem. Good performance then relies on this optimisation being as robust and efficient as possible. To aid in the speed and reliability of this convergence, we use three tricks in the course of this optimisation:

\begin{enumerate}
    \item Using point estimators of the model parameters for a good \sqm{seed} location for the optimisation.
    \item Using the optimum of each slice as the start for the following slice, and ordering these to be as smooth as possible, and
    \item Preconditioning the function to be optimised to make the posterior as close as possible to a unit Gaussian.
\end{enumerate}

\subsection*{Estimating Seed Parameters}
To initialise the Laplace and SVI quadrature methods and the grid generation, we attempt to find the maximum a posteriori estimate (MAP), i.e. the highest point in the posterior density. This optimisation requires an initial \sqm{seed} location. If not specified, this is calculated assuming stationary signal statistics to estimate the mean and amplitude of the continuum and response signals, i.e.:
\begin{equation}    
\begin{aligned}
    \mu_{\mathrm{cont/resp}}
    &= \left(\sum_i E_i^{-2}\right)^{-1} 
    \sum_i{E_i^{-2} y_i},
    \\
    \sigma_{\mathrm{cont/resp}}
    &= \left(\sum_i E_i^{-2}\right)^{-1}
    \sum_i{E_i^{-2}(y_i-\mu)^2}.
\end{aligned}
\end{equation}

To find a good starting value for $\tau$, we use the ICCF method outlined in Section~\ref{sec: existing_methods-nonparametric} to estimate the auto-covariance function of the continuum. This estimate becomes increasingly less accurate at high autocorrelation times $\delta t$, but drawing some window $\lvert \delta t\rvert<a$ it should follow close to the exponential form in Equation~\ref{eq: DRW_covar}.  Knowing this, we can say that $\logof{\mathrm{ACF}(\delta t)}\approx- \lvert \delta  t \rvert /\tau+2 \logof{\sigma}$. Drawing some window $\lvert \delta t\rvert<a$ and inverting the $\delta t>0$ side, this logarithm becomes roughly linear with slope $\frac{\partial y}{\partial \delta t}$. Taking a standard linear regression of this function, we can then estimate $\tau$. Precision is not too important here, as the model parameter is $\logof{\tau}$ in GP fitting and so is insensitive to small deviations.

\subsection*{Moving from Slice to Slice}

Under the assumption that the optimum is a smooth function of lag (see Figure~\ref{fig: litmus_slices} for an illustration), we use the solution for each slice as the starting location for the next slice. This is only done if this solution is accepted as being a valid representative of the marginal posterior's local behaviour. Solutions are rejected if they:
\begin{enumerate}
    \item Appear to be diverging, i.e. have undefined computation results, or
    \item Do not have a positive definite covariance matrix (i.e. the Laplace approximation fails), or
    \item Exhibit a severe drop in the peak posterior density compared to the previous slice, which would suggest the slice is in one of the furrows shown in Figure~\ref{fig: aliasing_explained}.
\end{enumerate}

If an optimisation diverges, a second attempt is made by resetting the start location to the seed parameters. If this second attempt fails, the slice is discarded for all further calculations / processing. 

\subsection*{Preconditioning}
For finding the peak density, \litmus uses the BFGS algorithm. This algorithm performs best in unit-quadratic functions, and we gain significant improvement in the convergence rate by linearly preconditioning the density function to warp it into this shape, i.e. rather than optimising $\logof{P_i( \phi)}$ we optimise $\logof{P_i( \phi(x))}, \; x = A(\phi-\phi_0)$, where $\phi_0$ is an a priori estimate of $\hat{\phi_i}$ and the starting location for optimisation, and $A$ is a preconditioning matrix encoding the local principle axes and their widths. Assuming $P_i(\phi)$ is roughly Gaussian, then we can construct suitable $A$'s from the Hessian evaluated at $\phi_0$ in three ways:

\begin{enumerate}
    \item From the Cholesky decomposition of the inverse Hessian, $AA^T=H^{-1}$.
    \item Performing a PDP decomposition of $H$ to find the covariance axes and corresponding Gaussian widths, $A=PD^{1/2}P$.
    \item An approximate form of (2) where we take only the diagonal elements of $H$, i.e. no skewing / rotating of the axes.
\end{enumerate}

For stability purposes, any negative or zero eigenvalues are fixed to unity in approaches (2) and (3).

\subsection*{Convergence \& Uncertainty Estimates}
\label{app: litmus-convergencetests}

Though the \hescan's approximation of the slice distribution log-densities as quadratic functions means that its convergent value is not exactly equal to the true model evidence, it is still necessary to constrain the deviation of our numerical calculation from this convergence point. There are two sources of numerical uncertainty in the calculation of $Z$: convergence of the optimiser to the true peak and the truncation error in the  integral.

In Equation~\ref{eq: Laplace_log_evidence}, errors in $\logof{Z_i}$ are determined to first order by the estimate of the peak density $\logof{P_i(\hat{\phi}^i)}$. Under a simple Newtonian optimisation, the next required \sqm{uphill} step is $\Delta \phi = H^{-1} \nabla f$, and the associated uncertainty in the peak density is $\nabla f^T \Delta \phi$, i.e.:
\begin{equation}
    \Delta \logof{Z_i} = - \nabla f^TH^{-1} \nabla f
    .
    \label{eq: hessian_tol_uncert}
\end{equation}
Conveniently, this also has a natural interpretation in terms of the \qm{closeness to the peak}. Noting that the Hessian is the negative inverse of the covariance matrix, this may also be written $ - \Delta \phi^T H  \Delta \phi=\Delta \phi ^T C^{-1}\Delta \phi$, i.e. the square \sqm{number of standard deviations} that the estimated solution is from the peak under the Gaussian approximation. This gives both a measure of the numerical uncertainty in $\logof{Z_i}$ and also a natural measure of \qm{closeness} to the optimum.

The second source of numerical uncertainty is the integration error due to the finite number of slices in the quadrature. To estimate this, we treat the integral as having $E_\mathrm{int} \propto\mathcal{O}(h^2)$ error scaling and sub-sample the slices, using only every second value to get a second less precise estimate $Z_{1/2} \pm 4\times E_\mathrm{int}$. Approximating the two estimates as having uncorrelated errors, their difference is ${Z-Z_{1/2}=0  \pm \sqrt{E_\mathrm{int}^2+(4\times E_\mathrm{int}^2)}}$, from which we can estimate the integral uncertainty $E$ as:
\begin{equation}
    E_\mathrm{int} \approx \frac{1}{\sqrt{17}}\lvert Z-Z_{1/2}\rvert
    \label{eq: integ_uncert}
\end{equation}
The convergence uncertainty always under-estimates $Z_i$ at each slice, $\Delta \logof{Z_i}<0$, and so we assume a worst case sum for the total convergence uncertainty. These two sources are then added in quadrature for a total evidence error estimate:
\begin{equation}
    \Delta Z^2 = E_\mathrm{int}^2 + 
    \left(\sum_i{Z_i \times\exp({\Delta \logof{Z_i})}} \right)^2
    .
    \label{eq: int_uncert}
\end{equation}
For the uncertainty in the \sviscan method, we use Equation~\ref{eq: int_uncert} but with the $\Delta \logof{Z_i}$ of Equation~\ref{eq: hessian_tol_uncert} instead estimated from the standard deviation of the ELBO over multiple SVI iterations assuming a $1/\sqrt{N}$ scaling of the uncertainty.

It is worth emphasising that this represents only the numerical uncertainty arising from convergence of the \hescan and \sviscan algorithms, not the uncertainty in the evidence that arises from the Gaussian approximation of the posterior density.

\subsection*{Computational Speed \& Tuning}
\label{app: speed_and_tuning}

A major feature of \litmus is that it offers complete Bayesian integrals without exorbitant computational cost. As a part of the scale-testing in Section~\ref{sec: validation}, we measure also the full runtime of the different algorithms, inclusive of null hypothesis model fitting for those methods that can do evidence integrals. The examples here are deliberately under-tuned, with default values for most fitting parameters (see Table~\ref{tab: fitting_params}) but with the number of iterations in the \sviscan ELBO calculations at each slice deliberately set high to ensure convergence.

\begin{figure}
    \centering
    \includegraphics[width=1.0\linewidth]{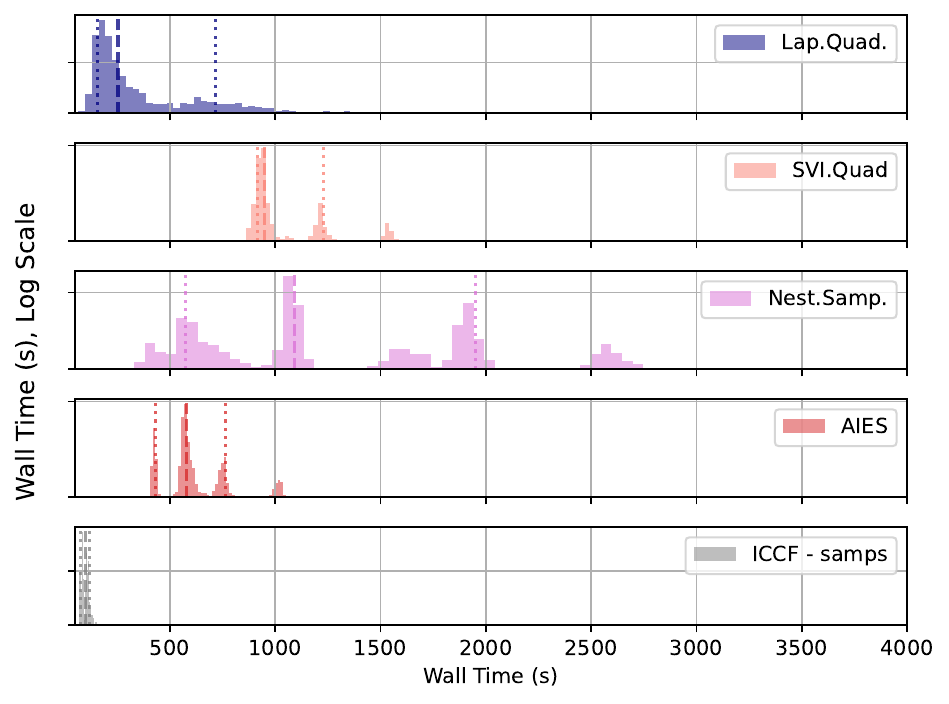}
    \caption{Histogram of the run-times for the five fitting methods in \litmus over all mocks, using the fitting parameters described in Table~\ref{tab: fitting_params}. The ICCF method, which requires no matrix inversion owing to its absence of GP fitting, is consistently the fastest. The \hescan can perform very fast except for cases where it get stuck optimising at a new test lag when the local optimum changes quickly over the lag axis. The \sviscan has a similar issue, but runs overall somewhat slower.}
    \label{fig: runtime_histogram}
\end{figure}

Shown in Figure~C-\ref{fig: runtime_histogram}, we can see that \litmus's various fitting methods are of comparable speed to the \javelin-like AEIS with default parameters while still being significantly more reliable (see Section~\ref{sec: validation}). The multimodality in these distributions comes from periods in the fitting of mocks with inconvenient posteriors during which proposals / new live points are difficult to find. 

It is difficult to fairly compare the run times of different algorithms due to the long list of tuning parameters for each method. With proper tuning, the \sviscan and \hescan algorithms can run markedly faster. Note also that this is not a direct comparison with \javelin, only with \litmus's example implementation of its fitting method, which is still built in \jax and \numpyro. In this way Figure~C-\ref{fig: runtime_histogram} is a comparison of the speed of the algorithms and not the codes themselves. 

\begin{table}[]
    \footnotesize
    \centering
    \begin{tabular}{|c|c|}
        \hline\hline
        \multicolumn{2}{|c|}{\hescan Parameters} \\\hline
        Number of Test lags & 128\\\hline
        Grid Bunching Parameter & 0.5\\\hline
        Slice Optimisation Tolerance & 0.01\\\hline
        Slice Optimisation BFGS Step Size & 0.001\\\hline
        Max Optimisation Steps & 1000\\\hline
        Slice Discard Log-Density Threshold & 10000\\\hline
        \hline
        \multicolumn{2}{|c|}{\sviscan Parameters} \\\hline
        Grid Bunching Parameter & 0.5\\\hline
        SVI ELBO Optimisation Steps & 1000\\\hline
        SVI Slice Discard ELBO Threshold & 1000\\\hline
        ELBO adam Step Size & 0.005\\\hline
        Number of ELBO Particles & 128\\\hline
        \hline
        \multicolumn{2}{|c|}{Nest. Samp. Parameters} \\\hline
        Number of live points & 1000\\\hline
        Max samples & 10000\\\hline
        Target evidence uncertainty & 0.001\\\hline
        \hline
        \multicolumn{2}{|c|}{AEIS Parameters} \\\hline
        Number of Walkers in Ensemble & 256\\\hline
        Total Number of Samples & 200000\\\hline
        Warmup Samples & 5000\\\hline
        \multicolumn{2}{|c|}{ICCF Parameters} \\\hline
        Number of Lags in Grid & 512\\\hline
        Number of Bootstraps & 512\\\hline
        Number of Times for Light Curve Interpolation & 1024\\\hline
    \end{tabular}
    \caption{Tuning parameters for the fitting methods used in Section~\ref{sec: validation}.}
    \label{tab: fitting_params}
\end{table}

\twocolumn
\bibliographystyle{apj}
\bibliography{bib_main,bib_Hbetasources,bib_MgIIsources,bib_CIVsources,bib_OzDES,bib_LITMUS}

\end{document}